\documentclass{aa}
\usepackage[varg]{txfonts}

\usepackage{natbib}
\usepackage{xspace}
\usepackage{siunitx} % \num{2.88(32)E-3}
\usepackage{graphicx}
\usepackage{xcolor}

\usepackage{hyperref} %always as the last package! footnote warnings...
\newcommand*\ie{i.e.,\ } %{id est\xspace} %{i.e.\ }
\newcommand*\eg{e.g.,\ } % {exempli gratia\xspace} % {e.g.\ }
\newcommand*\dd{\textnormal{d}}

\newcommand*\erosita{eROSITA\xspace}
\newcommand*\chandra{{\it Chandra}\xspace}
\newcommand*\xmm{{\it XMM-Newton}\xspace}
\newcommand*\rosat{ROSAT\xspace}
\newcommand*\sixte{SIXTE\xspace}
\newcommand*\esass{eSASS\xspace}
\newcommand*\cscale{\SIrange[range-units = single, range-phrase=--]{0}{1}{\arcmin}\xspace}
\newcommand*\sscale{\SIrange[range-units = single, range-phrase=--]{1}{4}{\arcmin}\xspace}
\newcommand*\lscale{\SIrange[range-units = single, range-phrase=--]{1}{16}{\arcmin}\xspace}

\hypersetup{
colorlinks=true,        % false: boxed links; true:colored links
linkcolor=black,        % color of internal links(change box color with linkbordercolor)
citecolor=blue,         % color of links to bibliography
filecolor=black,        % color of file links
urlcolor=cyan           % colour of url links
                }

\graphicspath{
{./results/}
}

\begin{document}

%increase the row height in a table
\renewcommand{\arraystretch}{1.5}

\title{Toward the low-scatter selection of X-ray clusters} 
\subtitle{Galaxy cluster detection with eROSITA through cluster outskirts} 

\author{
Florian K{\"a}fer\inst{1}
\and Alexis Finoguenov\inst{2}
\and Dominique Eckert\inst{3}
\and Nicolas Clerc\inst{4}
\and Miriam E. Ramos-Ceja\inst{1}
\and Jeremy S. Sanders\inst{1}
\and Vittorio Ghirardini\inst{1,5}
} 

\offprints{F. K{\"a}fer, \email{fkaefer@mpe.mpg.de}}

\institute{
Max-Planck-Institut f{\"u}r extraterrestrische Physik, Giessenbachstra{\ss{}}e, 85748 Garching, Germany
\and Department of Physics, University of Helsinki, PO Box 64, 00014, Helsinki, Finland
\and Department of Astronomy, University of Geneva, Ch. d'Ecogia 16, 1290 Versoix, Switzerland
\and IRAP, Universit\'e de Toulouse, CNRS, UPS, CNES, Toulouse, France
\and Harvard-Smithsonian Center for Astrophysics, 60 Garden Street, Cambridge, MA 02138, USA
}

\date{Received / Accepted }

\abstract
{%Context
One key ingredient in using galaxy clusters as a precision cosmological probe in large X-ray surveys is understanding selection effects. The dependence of the X-ray emission on the square of the gas density leads to a predominant role of cool cores in the detection of galaxy clusters. The contribution of cool cores to the X-ray luminosity does not scale with cluster mass and cosmology and therefore affects the use of X-ray clusters in producing cosmological constraints.
}
{%Aims
One of the main science goals of the extended ROentgen Survey with an Imaging Telescope Array (\erosita) mission is to constrain cosmology with a wide X-ray survey. We propose an \erosita galaxy cluster detection scheme that avoids the use of X-ray cluster centers in detection. We calculate theoretical expectations and characterize the performance of this scheme by simulations. 
}
{%Methods
We performed Monte Carlo simulations of the upcoming \erosita mission, including known foreground and background components. By performing realistic simulations of point sources in survey mode, we searched for spatial scales where the extended signal is not contaminated by the point-source flux. We derive a combination of scales and thresholds, which result in a clean extended source catalog. We designed the output of the cluster detection, which enables calibrating the core-excised luminosity using external mass measurements. We provide a way to incorporate the results of this calibration in producing the final core-excised luminosity.  
}
{%Results
Similarly to other galaxy cluster detection pipelines, we sample the detection space of the flux -- cluster core radius of our method and find many similarities with the pipeline used in the 400d survey. Both detection methods require large statistics on compact clusters in order to reduce the contamination from point sources. The benefit of our pipeline consists of the sensitivity to the outer cluster shapes, which are characterized by large core sizes with little cluster to cluster variation at a fixed total mass of the cluster.
}
{%Conclusions
Galaxy cluster detection through cluster outskirts improves the cluster characterization using \erosita survey data and is expected to yield well-characterized cluster catalogs with simple selection functions.
} 


\maketitle

%
%%
%%% Section 1
\section{Introduction}
\label{sec:introduction}

The expansion and structure formation history of the Universe is imprinted on the spatial distribution and number density of its largest collapsed entities, galaxy clusters. This makes galaxy clusters powerful probes for constraining cosmological parameters such as the dark energy equation of state (\eg \citealt{2009ApJ...692.1060V}; \citealt{2011ARA&A..49..409A} for a review). Among others, X-ray observations of galaxy clusters are of particular interest because they trace the bulk of the baryonic component, the hot intracluster medium (ICM). With the launch of the extended ROentgen Survey with an Imaging Telescope Array (\erosita, \citealt{2012arXiv1209.3114M,2018SPIE10699E..5HP}) in July 2019, X-ray astronomy ushers in a new era. As the primary instrument of the Russian-German Spektrum-Roentgen-Gamma (SRG) mission, \erosita will perform eight all-sky surveys within four years. The unprecedented survey speed and capability over a wide range of energies mean that the final all-sky survey will be $\sim$\numrange[range-phrase=--]{20}{30} times deeper than that of its predecessor (the \rosat all-sky survey, \citealt{1999A&A...349..389V}) in the \SIrange[range-units = single, range-phrase=--]{0.5}{2}{\keV} energy range and will provide the first ever imaging all-sky survey in the \SIrange[range-units = single, range-phrase=--]{2}{10}{\keV} energy band. With the expected detection of \num{e5} galaxy clusters \citep{2012MNRAS.422...44P}, \erosita will place tight constraints on the dark energy equation of state, among others.

Understanding selection effects is an essential but complicated requirement for precision cosmology. Determining the selection function is especially complex for extended X-ray sources because the detection probability and proper classification depend on their morphology, for example \citep{2011A&A...526A..79E,2016MNRAS.457.4515R,2017MNRAS.468.1917R,2017ApJ...843...76A,2017ApJ...846...51L}. The cluster outskirts (\numrange[range-phrase=--]{0.2}{0.8}\,$r_{500}$) are found to evolve with redshift in a self-similar fashion \citep{2017ApJ...843...28M,2019A&A...628A..43K} and exhibit low scatter \citep{2018A&A...614A...7G,2019A&A...628A..43K}. Therefore, cluster samples that are selected based on the properties of cluster outskirts will closely trace the selection by cluster mass and reduce the systematics of cluster use in cosmological studies. 
Another important aspect of detailed image decomposition consists of the removal of point sources. In the extragalactic sky, the X-ray point-source population is dominated by active galactic nuclei (AGN). Active galactic nuclei cause false detections through the noise in the realization of their photon distribution. In addition, they contribute to the total flux of the cluster because the AGN halo occupation distribution extends to high masses, especially at high redshifts \citep{2012ApJ...758...47A,2014ApJ...790...43O}. The importance of AGN in contaminating cluster fluxes of \erosita observations has been highlighted by \citet{2018MNRAS.481.2213B}. 

Spatial filtering of X-ray images to describe the emission that is produced on different spatial scales has been introduced by \citet{1991ESOC...38...17S} and was successfully applied for source detection in cluster cosmology \citep{1998ApJ...502..558V,2006MNRAS.372..578P}.
%\LEt{to properly include these references into the main text, please substitute commas for the semicolons and add "and" before the last reference. This is a LaTeX command error that I cannot fix for you with the program I work with (citet and citep)}
\citet{2009ApJ...704..564F,2010MNRAS.403.2063F,2015A&A...576A.130F}, \citet{2013ApJ...765..117E}, \citet{2015ApJ...799...60M}, and \citet{2019MNRAS.483.3545G} applied the method to detect groups and clusters of galaxies using only the large scales of the X-ray emission. In this paper, we present the adaptation of the wavelet decomposition method for \erosita. 

The paper contains the characterization of the \erosita point-spread function, simulations of \erosita observations of the extragalactic fields, calibration of the point-source model, description of the cluster detection pipeline, and its characterization using synthetic simulations. 

Throughout this paper we assume a WMAP9 cosmology with a matter density, vacuum energy density, and Hubble constant of $\Omega_{\textnormal{m}} = \num{0.282}$, $\Omega_{\Lambda} = \num{0.719}$, and $H_{0} = \SI{69.7}{km\,s^{-1}\,Mpc^{-1}}$, respectively \citep{2013ApJS..208...19H}. The dimensionless Hubble function is defined as $E(z) = \sqrt{\Omega_{\textnormal{m}}(1+z)^3 + (1-\Omega_{\textnormal{m}}-\Omega_{\Lambda})(1+z)^2 + \Omega_{\Lambda}}$. Quoted errors are $1\sigma$ unless otherwise stated.

%
%%
%%% Section 2
\section{eROSITA and the eROSITA simulator}
\label{sec:simulation}

\erosita is a new X-ray telescope that was launched in July 2019 on board the SRG. The full description of the telescope can be found in \citet{2018SPIE10699E..5HP}.

\subsection{Point-spread function}
\label{subsec:psf}

The point-spread function (PSF) of an X-ray telescope describes its ability to focus photons. The image produced by a point source is blurred, mostly as a result of misalignments and micro-roughnesses or is caused by the support structures of the instrument's grazing incidence mirrors. The shape and size of the PSF depends among others on the photon energy and its distance from the optical axis. The current \erosita PSF model is based on measurements made at the PANTER X-ray test facility, where the PSF is sampled on an \num{11 x 11} grid, plus an additional central \num{6 x 6} grid to increase the small off-axis angle density. Each grid is spaced by \SI{6}{\arcmin} , and the two grids are displaced by \SI{3}{\arcmin} with respect to one another. The energy dependence is sampled using X-ray emission lines at photon energies of \SIlist[list-units = single,list-final-separator = {, and }]{0.3;0.9;1.5;3.0;4.5;6.4;8.0}{\keV}. The PSF image at each position and energy is described by shapelets \citep{2003MNRAS.338...35R}, that is, by a linear image decomposition into a series of differently shaped basis functions of characteristic scales. The shapelet description is a convenient way to compress the PSF information over a few coefficients. Two different scale parameters with individual shapelet coefficients are used in order to reproduce the complex behavior of the PSF core on small scales and the PSF wings on large scales. Each of the seven \erosita mirror modules are made out of 54 nested Wolter-I type \citep{1952AnP...445...94W} shells and have their individual PSF measurements. However, in the current implementation of the X-ray telescope simulator (Sect. \ref{subsec:simulator}), the PSFs of all seven modules are assumed to be the same, using only the shapelet reconstruction of flight module number \num{2}. We note that the \erosita PSF will be different in orbit, for example, due to shaking of the telescope during launch or temperature and gravitational effects. During the performance-verification and all-sky survey phases, the \erosita PSF will be determined and calibrated against ground-based measurements.

\subsection{Point sources and background components}

We followed the recipe of \citet{2018A&A...617A..92C} and used \sixte\footnote{\url{http://www.sternwarte.uni-erlangen.de/research/sixte/}} \citep[][see Sect. \ref{subsec:simulator}]{2019arXiv190800781D} to simulate \erosita fields containing AGN and unresolved X-ray background. Individual AGNs were drawn from a luminosity function down to a field exposure time-dependent flux threshold and uniformly distributed in a field. Thus spatial clustering of AGNs and spatial correlations between AGNs and galaxy clusters are not considered; this is the topic of a future study. The AGN spectra of the low-luminosity tail of the distribution were stacked and redistributed uniformly to construct an unresolved X-ray background component. Emission of the hot plasma in the halo and disk of our Galaxy was simulated using a double MEKAL model \citep{1985A&AS...62..197M,1986A&AS...65..511M,1995ApJ...438L.115L} with temperatures of \SI{0.081}{\keV} and \SI{0.204}{\keV} \citep{2002A&A...389...93L}. In addition, a non-vignetted \erosita instrument particle background component according to the expected radiation level at the Lagrange point L2 was simulated \citep{2010SPIE.7742E..0YT}.

\subsection{Extended objects}
\label{subsec:extobjects}

We here focus on the detection of extended sources. To compare our results to previous studies, we characterize the spatial flux distributions by spherically symmetric $\beta$-models \citep{1978A&A....70..677C} with $\beta=2/3$ on a discrete grid of core radii. The cluster emission was characterized by a partially absorbed Astrophysical Plasma Emission Code \citep[APEC,][]{2000HEAD....5.2701B} model with a fixed abundance of \SI{0.3}{\ensuremath{\textnormal{Z}_{\odot}}} \citep{1989GeCoA..53..197A} and a survey-field-dependent Galactic column density of hydrogen. The Galactic absorption was described by a phabs model \citep{1992ApJ...400..699B} and was fixed to \SI{3e20}{\per\cm\squared}, \SI{8.8e20}{\per\cm\squared}, and \SI{6.3e20}{\per\cm\squared} for the equatorial, intermediate, and deep field, respectively (see Sect. \ref{subsec:missionplanning}). Cluster temperatures, redshifts, and fluxes were sampled on a grid and ranged between
\SIrange[range-units = single, range-phrase=--]{1}{5}{\keV},
\numrange[range-phrase=--]{0.05}{1.2}, and
\SIrange[range-units = single,range-phrase = --]{2e-15}{5e-15}{erg\,s^{-1}\,cm^{-2}}, respectively.

\subsection{X-ray telescope simulator}
\label{subsec:simulator}

The simulations of the extragalactic \erosita sky were performed using the Monte Carlo based \sixte simulator \citep{2019arXiv190800781D}. A sample of photons was produced based on the effective area of the instrument and input source characteristics, for example, a source spectrum, or if necessary, a model of the extent. These photons were virtually propagated through the instrument simulator. Based on the telescope specifications, a list of impact times, positions, and energies of the photons was produced. The final output event list was then created by simulating the read-out characteristics. The simulator already provides an implementation of the \erosita characteristics described by the PSF, vignetting, response matrix files, and ancillary response files.

\subsection{\erosita mission planning and survey fields}
\label{subsec:missionplanning}

We assumed a simple survey strategy for the four-year all-sky survey, where the scanning axis is pointed toward the Sun and \erosita scans one great circle every four hours \citep{2012arXiv1209.3114M}. One full coverage of the sky is achieved every half year. We note that the final survey strategy will be more complicated due to additional constraints. Since the attitude file we used was created, the movable antenna was replaced by a fixed antenna, thus the spacecraft needs to perform compensating motions to maintain the angular constraints with respect to the Earth and the Sun. In addition, the antenna opening angle and the spacecraft-Sun-vector constraints were changed. This leads to a more inhomogeneous exposure in ecliptic longitude, among others.

We studied three \SI{3.6 x 3.6}{\degree} sky tiles with approximately \SI{2}{\kilo\second}, \SI{4}{\kilo\second}, and \SI{10}{\kilo\second} exposure. We refer to these fields as equatorial, intermediate, and deep, respectively. Taking vignetting into account, the median net exposures of the fields were roughly halved, that is, approximately \SI{1}{\kilo\second}, \SI{2.5}{\kilo\second}, and \SI{6}{\kilo\second}, respectively. The equatorial field shows a uniform exposure, but the deep field has a large exposure gradient \citep{2018A&A...617A..92C}.

%
%%
%%% Section 3
\section{Source detection and characterization}
\label{sec:detection}

The standard technique when source catalogs are created is to split source detection and characterization because different optimized software packages perform better on the individual tasks. After the initial detection, a maximum likelihood (ML) source characterization is used to separate extended and point-like sources, based on the value and the significance of the extent \citep{1998ApJ...502..558V,2007ApJS..172..561B,2016A&A...592A...2P,2018A&A...617A..92C}.
The approach of splitting detection and characterization is also implemented in the standard \erosita data-processing pipeline based on the \erosita Science Analysis Software System (\esass)\footnote{\url{https://erosita.mpe.mpg.de/eROdoc/}}.
The forward-fitting routine employed by the ML fitting ensures the best sensitivity toward detecting an object with the assumed characteristics. However, the assumed symmetric $\beta=2/3$ model is too simplistic for many extended sources.
The goal of our investigation is to provide a framework that selects extended sources based on their extended emission rather than relying on a blind fitting method.
Our galaxy cluster detection scheme is physically motivated and sensitive to the outer self-similar cluster regions. This ensures cluster selection from the point of view of best cluster characterization because the outer cluster regions show less scatter at a given cluster mass.

\subsection{Wavelet decomposition method}
\label{subsec:wvdecompmethod}

The general idea of wavelet decomposition is the isolation of differently sized structures by convolving the input image with kernels of variable scales. Starting with the smallest scale, significant emission is subtracted before continuing on the next larger scale. This allows us to model point-source emission based upon their detection on scales that are unresolved or are the size of the PSF. The angular sizes of these scales depend on instrumental and observational characteristics and can vary from arcseconds for the \chandra observatory to arcminutes for \rosat all-sky-survey data. We refer to these small scales as point-like emission detection (PED) scales, and greater scales are labeled extended emission detection (EED) scales.
% spatial scales used for the detection of extended emission
The removal of point sources based on the PED-scale detection and a PSF model prior to running the wavelet decomposition on EED scales is a natural step within the philosophy of wavelet decomposition and was introduced by \citet{2009ApJ...704..564F}. Following this approach, the general concept of our algorithm is to detect point sources and extended sources separately.
An overview of the general steps of our procedure is as follows:
\begin{enumerate}
  %%%
  \item Calibration of the point-source model on point-source-only simulations by obtaining normalization coefficients of PSF templates on PED scales.
  %%%
  \item Extended source detection and characterization on realistic simulations.
  \begin{enumerate}
    \item Detection of point sources on PED scales.
    \item Model the predicted point-source emission using normalization coefficients and PSF templates.
    \item Check for residual signal over the background and point-source emission on EED scales.
    \item Catalog extended sources.
  \end{enumerate}
\end{enumerate}

\textbf{[1.]}
The first step is the calibration of point-source modeling, which means addressing which angular sizes the PED scales have in the particular science case. We only used the simulated image of point sources, which contains resolved and unresolved sources. The background level was determined by iterating the detection of point-source emission and excising point sources from the background estimates, as was done for \xmm and \chandra in \citet{2015A&A...576A.130F}. 
Next, we modeled the instrument PSF with a sum of Gaussians without assuming any prior knowledge about its shape. This has the advantage of being robust and fast to implement. We modeled the PSF up to a scale on which the emission is almost free of point-source contamination. These scales are defined as PED scales. Using the converged background estimate, we ran the detection of point sources {on the PED scales to obtain a wavelet image of resolved point sources. This point-source image} was smoothed with Gaussians of different widths to obtain fitting templates. We fit these templates to the wavelet-subtracted image to derive the amplitude of the image that best describes the residuals. We did this by cross-correlating the maps in order to take the covariance of the templates into account. The results are individual normalization coefficients for the used templates. These normalization coefficients were used to model the PSF effect in the simulations that contain extended sources.
We note that including actual PSF measurements might improve the description of the PSF wings, which cannot be characterized by our approach of combining several Gaussians.
The point-source subtraction technique has proven to be very efficient in deep X-ray fields and has also allowed the separation of extended sources due to inverse Compton scattering of the cosmic microwave background photons on the relativistic plasma of radio jets \citep{2010ApJ...715.1143F,2010MNRAS.409.1647J}.
The detection threshold is the level in the convolved image above which the peaks are statistically significant. For the purpose of subtracting or modeling point-source contamination, the detection thresholds considered extended to $3\sigma$ \citep{1998ApJ...502..558V}.

\textbf{[2.(a/b)]}
After the calibration of the point-source model, we ran the detection of point sources on the realistic simulations, which contain both point and extended sources. We smoothed the resulting point-source wavelet image with Gaussians of the same widths as in the calibration. These templates were multiplied by the normalization coefficients obtained in the calibration and were added to the unsmoothed wavelet image to model the point-source emission on the PED scales.

\textbf{[2.(c/d)]}
To preserve the Poisson statistics, we added the point-source model to the background estimate and searched for residual signal over the background plus point-source model to detect and catalog extended objects. As a result, we obtained maps that were free of point-source emission. The maps retain the spatial shape of the extended source emission, such that ellipticity can be measured, for example. In addition, the maps allow for a simple visual characterization of the detected emission. This can be a complicated task, for instance, if the cluster does not look like a $\beta$-model because of extended source confusion. Furthermore, maps obtained by different satellites can be combined, as was done for \chandra and \xmm observations in \citet{2015A&A...576A.130F}.
The choice of the detection thresholds for cataloging the extended sources depends on the objective, and they were adjusted to the desired level of completeness and purity of the catalog, as discussed in Sect. \ref{subsec:sfextsources}. Typically, the detection thresholds were at least 4$\sigma$.

The goal of our pipeline is to select sources based on the extended emission, compared to selecting sources based on a symmetric $\beta=2/3$ model fit to some angular range. Thus, our catalogs include sources with a greater variety of shapes. This detection scheme has obvious benefits at low-mass halos, such as galaxy groups, because they exhibit a wide variety of X-ray morphologies \citep[\eg][]{2006ApJ...646..143F,2007MNRAS.374..737F}. From the point of view of source selection, the effect of contaminating sources is very different between this pipeline and classical wavelets. Here, the ability to detect and select a cluster as an extended source might be reduced due to the large noise caused by point-source induced background, while in other methods, the source might be classified as a cluster because of the point-source contribution to the total flux.

If we were to only keep the emission above the selected detection threshold, we would discard the bulk of the source flux. Wavelets provide a secondary filtering threshold for estimating the region around the detected maximum where significant flux is detected. A lower filtering threshold compared to the detection threshold therefore minimizes the loss of source flux by keeping a larger region around the detected maximum. This region can be used in the flux estimation of the source in the point-source-subtracted map. However, setting the filtering threshold too low has the drawback of potentially including secondary peaks within the region around the main peak, which would normally not be detected. These secondary peaks might increase the number of spurious detections. Flux measurements within a wavelet reconstructed region have been extensively tested in \citet{2012ApJ...756..139C}.
Together with the source flux, the detection efficiency of galaxy clusters depends on their extent. To achieve a comparison to previous studies, we considered the performance of our pipeline by adopting the same framework as for $\beta$-model profiles. Within the $\beta$-model approach, the extent is characterized by the value of the core radius. A discussion to extend the existing $\beta$-model tools to capture the wide variety of expected source shapes is beyond the scope of this paper.

In addition to the standard $\beta$-model characterization, our pipeline can be calibrated using any set of cluster characterization. In addition, the catalog of extended sources can be fed back into the $\beta$-model-extent fitting routine to identify why certain sources are lacking from the cluster list. This approach is similar to the XMM-XXL survey pipeline \citep{2006MNRAS.372..578P}. We note that compared to our method, the flux estimate of the XMM-XXL pipeline includes potential excess cool-core emission.

\subsection{Adjusting the detection pipeline to eROSITA}

As described in Sect. \ref{subsec:wvdecompmethod}, the proposed source detection algorithm needs to be tuned to the characteristics of the particular observation. In this section, we focus on how to adapt the general framework to \erosita. Currently, our training is limited to the pre-flight calibration, and a further tuning of the pipeline is required in-flight. Compared to similar pipelines for \chandra and \xmm, we have not yet addressed the minor deficiencies associated with wavelet flux redistribution between adjacent scales. This will be accomplished as a part of the in-flight calibration and will serve to reduce the root mean square of the residual image. Right now, we propose an effective scheme of the procedure and apply it to current \erosita-survey mock observations. We follow this path because the incorporation of in-orbit PSF calibration data into the software analysis has proven to be time consuming in our experience.

Similar to \chandra and \xmm, the off-axis degradation of the \erosita PSF is driven by the fact that the detector plane is out of focus. Thus, we can directly apply our experience with developing the source detection algorithm for \chandra and \xmm to \erosita. However, the \erosita maximum degradation in terms of the half-energy width is 20\% \citep{2010SPIE.7732E..0UP}, which is lower than for \xmm and far lower compared to \chandra. The \erosita PSF does not have a core, but a typical survey half-energy width of $28^{\prime\prime}$ at \SI{1}{\keV}. In scanning mode, the \erosita PSF is roughly uniform across sky tiles.
The detector pixel size corresponds to $9.6^{\prime\prime}$ , and sky tiles are rebinned into images with $4^{\prime\prime}$ pixel size. In our simulations, the impact position of each photon is known.
In the \erosita survey, the rebinning will be made by reconstructing split events using the charge division among adjacent pixels, allowing for subpixel resolution \citep{2012SPIE.8443E..50D}.
We detect sources in \erosita-survey images in the \SIrange[range-units = single, range-phrase=--]{0.5}{2}{\keV} energy band. Events are not split or selected based on their off-axis angles.

\textbf{[1.]}
First, we study the limitations of the point-source-model process on the \erosita cluster detection. The goal is to answer the questions whether we can reliably model the point-source contribution and to define the angular scales required for this. The angular scales on which point sources are first detected is a strong function of the survey depths, instrument PSF, and assumed background. A discussion of the effects is presented in \citet{2015ApJ...799...60M}. With respect to \erosita survey observations, we are not able to reliably predict the residual point-source emission on scales below $32^{\prime\prime}$ because most of the point sources are only detected on the $32^{\prime\prime}$ wavelet scale. Even on scales of $64^{\prime\prime}$, we detect point sources that are not detected on any smaller scales. The point-source contamination on scales starting from $128^{\prime\prime}$ is minimal.
In training for the point-source model, we ran the wavelet decomposition up to a scale of $32^{\prime\prime}$. Because we are interested in a complete source subtraction, we adopted a low detection and filtering threshold of $3.3\sigma$ and $1\sigma$, respectively. These small scales are smoothed with Gaussians of $64^{\prime\prime}$ and $128^{\prime\prime}$ widths and fitted to the $32^{\prime\prime}$ wavelet-subtracted image. The two Gaussian-smoothed templates describe the residual image best, with normalization coefficients of 0.47 and 0.1, respectively. We did not include the $64^{\prime\prime}$ scale in modeling the point-source flux on the EED scales because we wished to retain sensitivity for extended objects on this scale.

\textbf{[2.(c/d)]}
The prediction of the point-source emission on PED scales was included in the background model, and we ran the wavelet decomposition on the EED scales in order to detect and catalog extended objects. A widely adopted way for cleaning catalogs is to set the detection threshold for extended sources higher, which reduces the chance of including misclassified point sources as extended. For \erosita, we did not detect point sources on the $64^{\prime\prime}$ scale when we set the detection threshold to $7\sigma$. On the other hand, scales starting from $128^{\prime\prime}$ are already very clean from the point-source contamination, and the lowest statistically motivated thresholds can be adopted there. This is very good news for the science of galaxy groups with \erosita, as well as for studies of the unresolved background fluctuation. For this work, we illustrate the performance of the pipeline using two detection thresholds: one maximally sensitive, of $4\sigma,$ and another maximally clean, of $7\sigma$. The filtering thresholds were set to 1.6$\sigma$ and 3$\sigma$ for high-sensitivity and low-contamination wavelet detection, respectively. In addition, we adapt a $5\sigma$ detection threshold with a 1.6$\sigma$ filtering threshold in Sect. \ref{subsec:sfextsources} for a better comparison with an existing study.
We motivate these thresholds further in Sect \ref{subsec:selection}.
Our current simulations do not consider spatial AGN clustering, and the quantification of this effect is topic of a future study. We note that a potential AGN clustering might create false fluctuations on larger scales.

Considering the shallow depths of the \erosita survey, the limitation of using EED scales is primarily for detecting sources at high redshift ($z\sim1$). There, using smaller scales for fitting the cluster shapes will be complicated by the enhanced AGN activity in clusters \citep{2018MNRAS.481.2213B} and might present a fundamental limitation of the survey to achieve clean high-$z$ cluster flux estimates in any case, as opposed to merely detecting a cluster.
When the in-orbit background is higher than we assume here, the detection threshold can be lowered, staying the same in terms of the source flux. Thus, our results in terms of source flux detection will be quite representative for a wide range of in-orbit conditions.

%
%%
%%% Section
\section{Selection criteria}
\label{sec:selcrit}

The point-source-cleaned maps provide a way to detect extended sources and to measure their flux. From the point of view of the flux extraction, it is clear that the flux on the spatial scales used to estimate the point-source flux will be partially removed. On the other scales, the work on \erosita sample construction has put forward a demand on defining the simplest possible observable on which the selection is made with a preferable step-function-like selection \citep{2018arXiv181010553G}. In our method, this is the residual cluster flux in the \sscale range.
This represents a simple aperture extraction, which is linked to the total cluster flux. The radial range is not directly motivated by the wavelet analysis, except that we need to consider scales above \SI{1}{\arcmin} due to point-source confusion (see Sect. \ref{subsec:sfextsources}). In addition, we also present a consideration of the source flux in the  \lscale range. This allows us to study the effects of using a larger aperture.
Our experiments with cluster detection in the equatorial fields led to a conclusion of using 40 and 80 counts for these two detection ranges, respectively. This assumes a lower detection threshold for the large area.
Previous studies \citep[\eg][]{2012MNRAS.422...44P,2014A&A...567A..65B} neglected aperture effects in addition to a count threshold and assumed a fixed minimum number of total photons to classify a source as cluster. This leads to an artificially high sensitivity toward a detection of X-ray emission from low-redshift galaxy groups. For cosmological studies, however, these systems are not the intended targets and can be neglected. In the following we derive the analytical description of the cluster selection based on these two thresholds.

It is clear that the redshift range for which our technique is the most attractive is also the range where the selected radial range samples the part of the cluster with the lowest scatter against the total mass. This corresponds to typical clusters of, for example, \SI{e14}{\ensuremath{\textnormal{M}_{\odot}}} at redshift 0.4. The sampled part of the cluster changes with redshift as well as mass, and we prefer to model this effect as opposed to changing the extraction region as a function of the redshift-dependent limiting mass. The actual reconstruction of the cluster properties does not have to follow this prescription, and several efforts are underway to provide the core-excised luminosity for the \erosita clusters (\eg Eckert et al. in prep.). 

Using the integrated counts (or count rates) is just one of the possibilities for cluster selection based on our maps. Our source lists can be used with the ML fitting in its standard form and in the modified form, in which the core radii of the clusters are examined only at large radii. This avoids the influence of the cool cores on the estimate, as found by \citet{2019A&A...628A..43K}. 

%
%%
%%% Section
\section{Theoretical predictions}
\label{sec:theo_prediction}
\begin{figure}
\centering
\resizebox{\hsize}{!}{\includegraphics{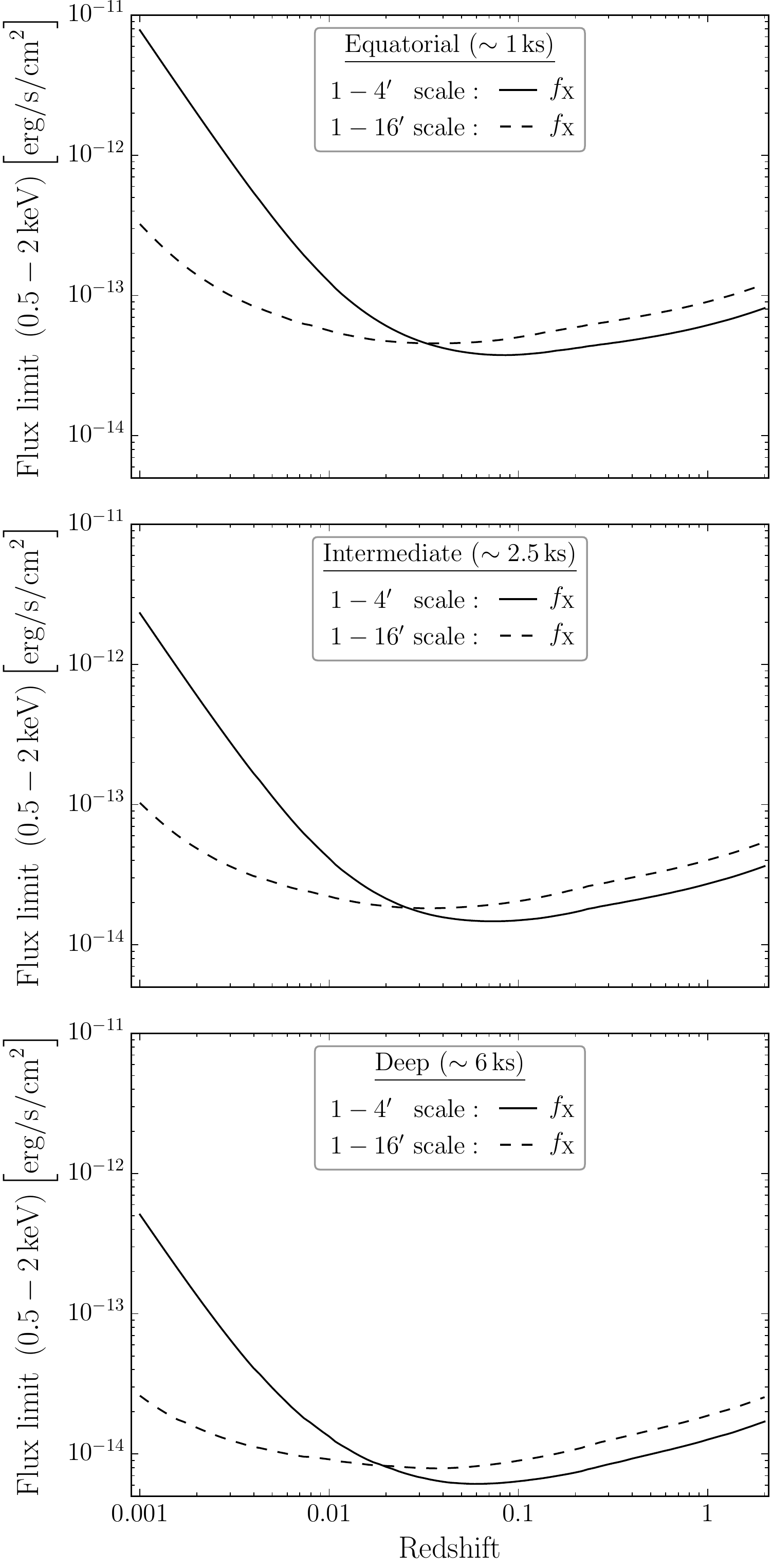}}
% For referee version; otherwise the figure is too big for the page...
%\resizebox{11cm}{!}{\includegraphics{flux_z_limit_all.pdf}}
\caption{Galaxy cluster flux limit as a function of redshift for an equatorial, an intermediate, and a deep final \erosita survey field of approximately \SI{1}{\kilo\second}, \SI{2.5}{\kilo\second}, and \SI{6}{\kilo\second} exposure, respectively. The black solid and black dashed lines show the flux limits corresponding to 40 and 80 counts in  the detection region of \sscale and \lscale radial scale, respectively.
}
\label{fig:flux_z_limit_all}
\end{figure}

\begin{figure}
\centering
\resizebox{\hsize}{!}{\includegraphics{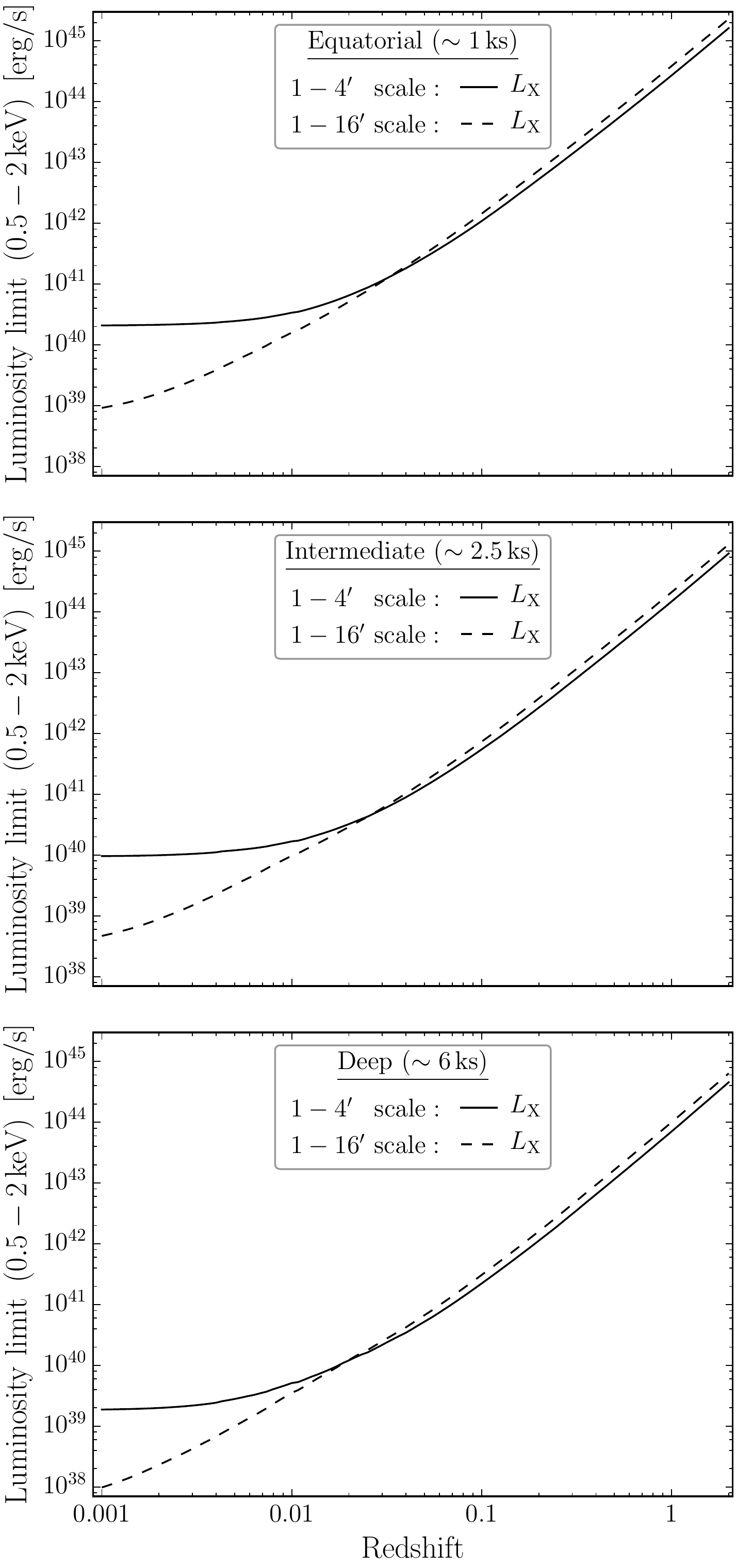}}
%\resizebox{11cm}{!}{\includegraphics{lum_z_limit_all.pdf}}
\caption{Same as Fig. \ref{fig:flux_z_limit_all} for the galaxy cluster luminosity limit.
}
\label{fig:lum_z_limit_all}
\end{figure}

\begin{figure}
\centering
\resizebox{\hsize}{!}{\includegraphics{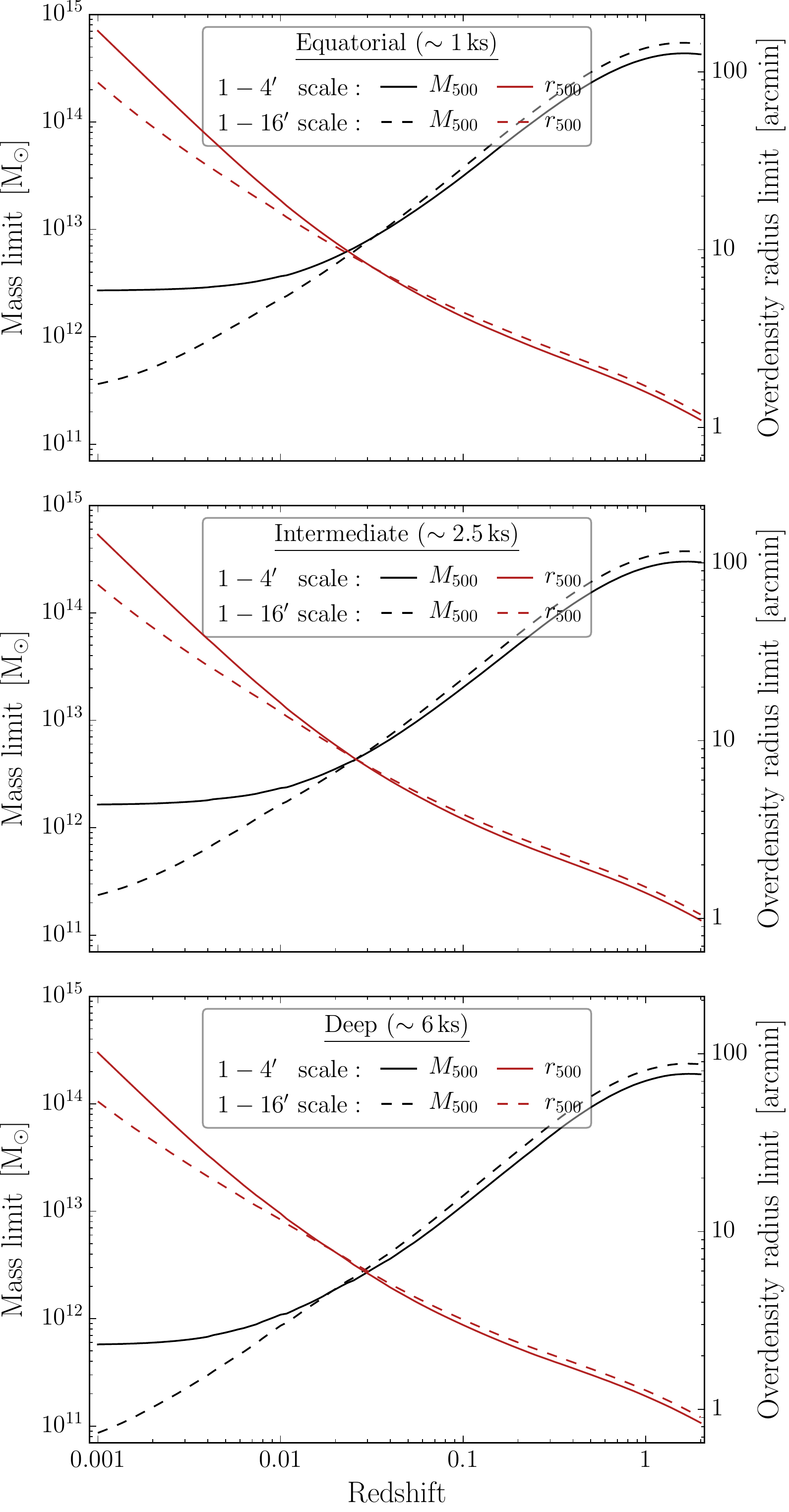}}
%\resizebox{11cm}{!}{\includegraphics{m500_z_limit_all.pdf}}
\caption{Same as Fig. \ref{fig:flux_z_limit_all} for the galaxy cluster mass limit. The brown solid and brown dashed lines represent the associated overdensity radii (right-hand $y$-axes).
}
\label{fig:m500_z_limit_all}
\end{figure}

\begin{figure}
\centering
\resizebox{\hsize}{!}{\includegraphics{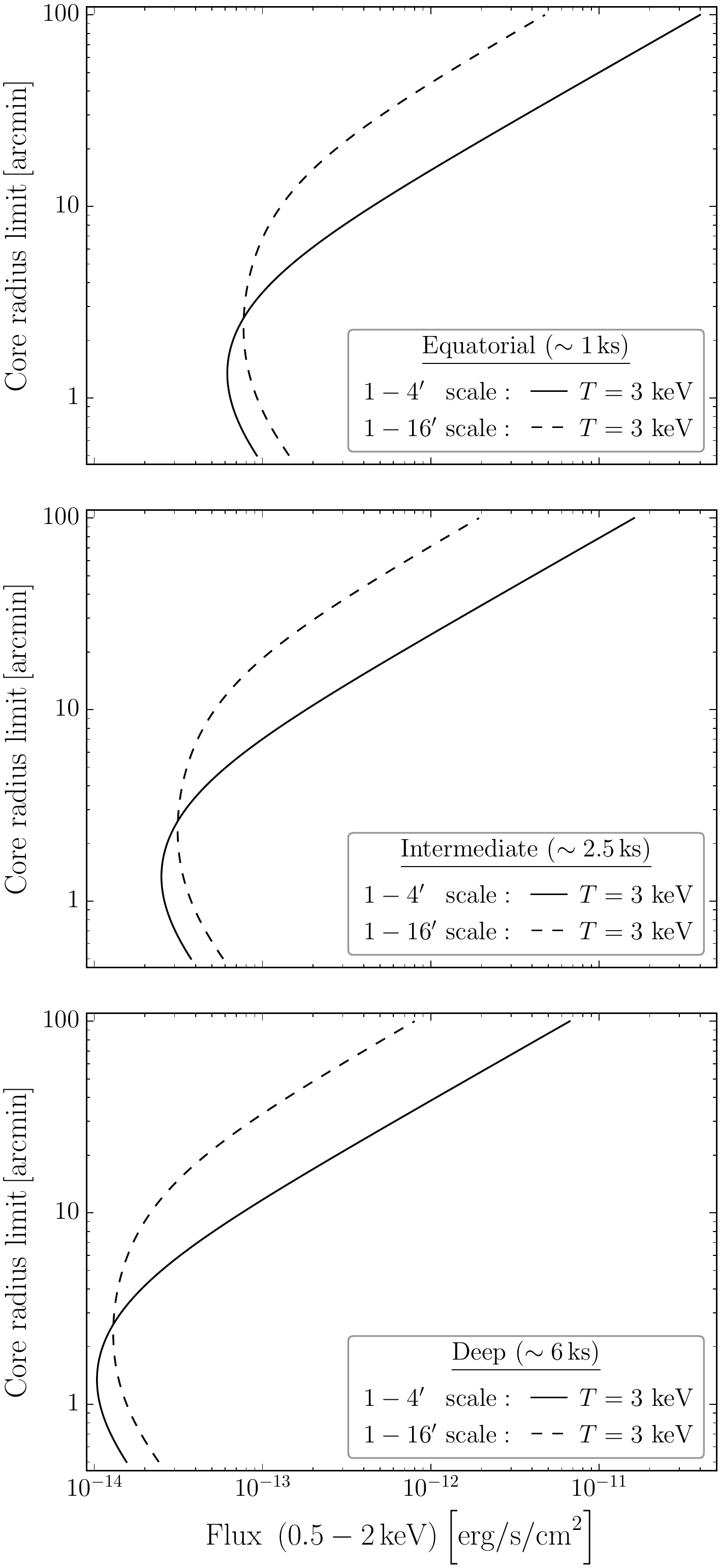}}
%\resizebox{10cm}{!}{\includegraphics{rc_flux_counts_limit_all.pdf}}
\caption{Galaxy cluster core radius limit as a function of flux for an equatorial, an intermediate, and a deep final \erosita survey field of approximately \SI{1}{\kilo\second}, \SI{2.5}{\kilo\second}, and \SI{6}{\kilo\second} exposure, respectively. The black solid and black dashed lines show the core radius limits of a \SI{3}{\keV} cluster corresponding to 40 and 80 $\beta$-model counts on a \sscale and \lscale radial scale, respectively.
}
\label{fig:rc_flux_counts_limit_all}
\end{figure}

\begin{figure}
\centering
\resizebox{\hsize}{!}{\includegraphics{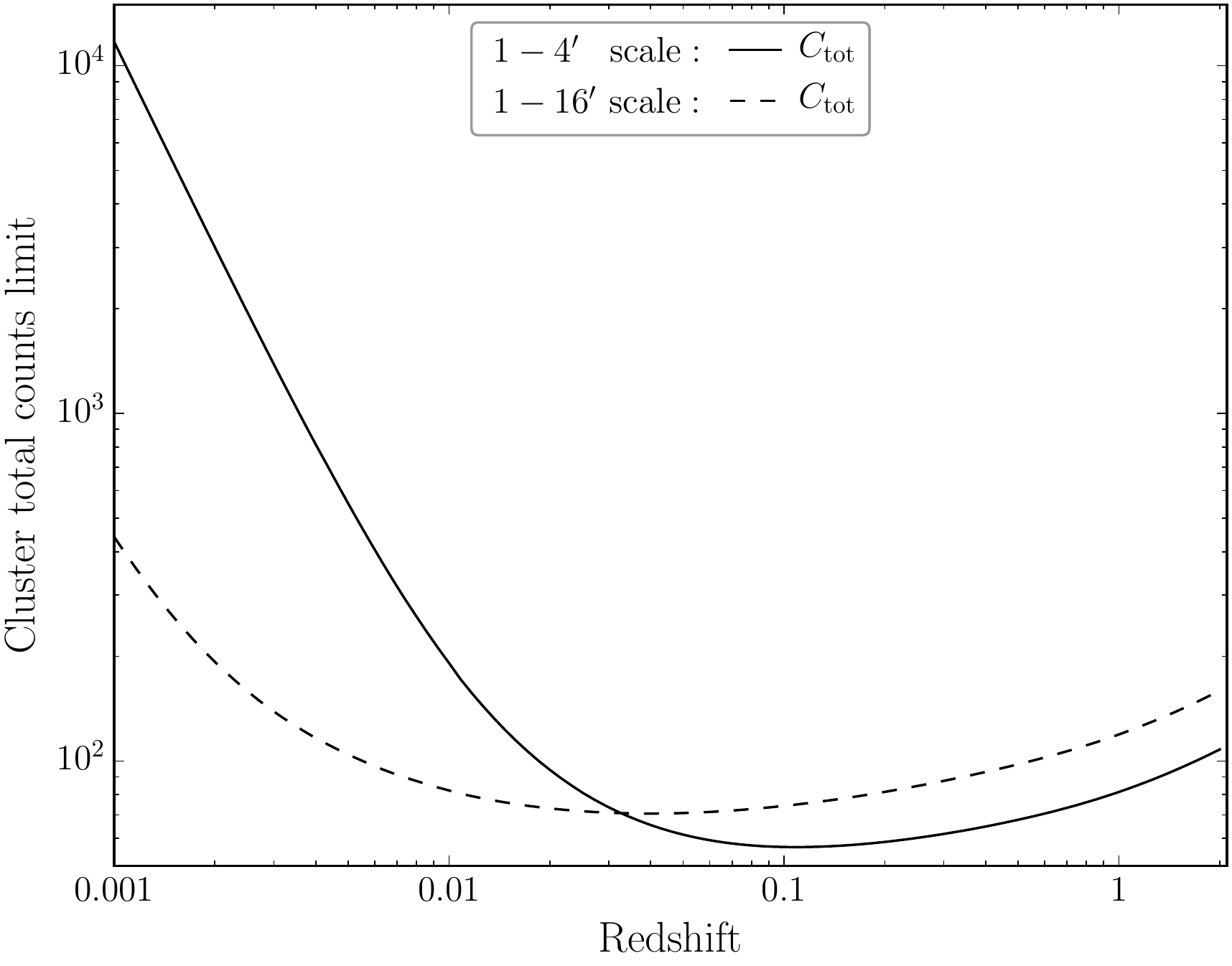}}
\caption{Galaxy cluster total count limit as a function of redshift. The black solid and black dashed lines show the total count limit, corresponding to 40 and 80 $\beta$-model counts on a \sscale and \lscale radial scale.
}
\label{fig:cntstot_z_limit}
\end{figure}

When we assume the minimum number of counts to detect a galaxy cluster as an extended object ($C_{\textnormal{det}}$) in a field with a given exposure time ($T_{\textnormal{exp}}$), we can iteratively calculate the corresponding cluster flux ($f_{500,\textnormal{lim}}$), luminosity  ($L_{500,\textnormal{lim}}$), and mass limit ($M_{500,\textnormal{lim}}$) as a function of redshift.
Given an initial cluster mass and temperature-limit guess, the corresponding overdensity radii are calculated assuming spherical symmetry through
\begin{equation}
r_{500,\textnormal{lim}} = \left( \frac{3 M_{500,\textnormal{lim}}} {4 \pi \cdot 500 \rho_{\textnormal{crit},z}} \right)^{1/3}.
\label{equ:r500}
\end{equation}
The core radii are assumed to scale with the overdensity radii ($r_{\textnormal{c}} = r_{500}/3$). This ensures that the apparent size scales with redshift, that is, clusters at higher redshift have a smaller angular extent. The relation between core and overdensity radii is calibrated on non-cool-core clusters at low redshift \citep{2019A&A...628A..43K} and holds at high redshift, where the relative contribution of the cool core to the outer parts of the cluster becomes minor \citep{2013ApJ...774...23M}. The compactness of clusters at high redshift matters for the detection. In practice, we need to characterize the detected population of groups and clusters and correct the numbers for the differential sensitivity of the detection method. With the core radius estimate, the count rate of a cluster is calculated by integrating a single $\beta$-model with fixed slope ($\beta=2/3$) in a given radial range. Realistic deviations from the $\beta=2/3$ assumption have little impact on the shown thresholds because the actual distribution of the counts is less important. We denote the $\beta$-model count rate on the \sscale scale  as $R(\SI{1}{\arcmin},\SI{4}{\arcmin})$ and the count rate within $r_{500}$ as $R(\SI{0}{\arcmin}, r_{500})$. Both predicted $\beta$-model count rates are independent of PSF redistribution effects. 
We used the X-ray spectral-fitting program XSPEC \citep{1996ASPC..101...17A} as well as the temperature guess to calculate the conversion factor of count rate to flux ($\lambda_{\textnormal{RF}}$) by dividing the model flux of a partially absorbed APEC model (see Sect. \ref{subsec:extobjects}) with unity normalization by the corresponding APEC-model count rate. The conversion factors of count rate to flux range between \SIrange[range-units = single, range-phrase=--,range-units = brackets, fixed-exponent = -13,scientific-notation = fixed]{6.45e-13}{7.65e-13}{erg\,count^{-1}\,cm^{-2}}.
The cluster flux limit is derived according to
\begin{equation}
f_{500,\textnormal{lim}} =  \lambda_{\textnormal{RF}}
\cdot
\frac{C_{\textnormal{det}}} {T_{\textnormal{exp}}}
\cdot
\frac{R(\SI{0}{\arcmin}, r_{500})} {R(\SI{1}{\arcmin},\SI{4}{\arcmin})}.
\label{equ:flulim}
\end{equation}
Using the redshift, we calculated the conversion factor of  count rate to luminosity ($\lambda_{\textnormal{RL}}$) by shifting the desired rest-frame energy band (\SIrange[range-units = single, range-phrase=--]{0.5}{2}{\keV}) to the observed one. This is in order to correct for the fact that the energies of detected photons in a given passband are ($1+z$) times lower than in the cluster rest-frame. The intrinsic APEC-model luminosity is calculated by multiplying the unabsorbed APEC-model flux in the observed band with $4\pi$ times luminosity distance squared and divided by the APEC-model count rate to obtain $\lambda_{\textnormal{RL}}$. 
The luminosity conversion factor as a function of redshift is roughly a broken power law, in the form of monomials, with break point around $z=0.1$. It ranges between \SIrange[range-units = single, range-phrase=--]{2e39}{2e43}{erg\,count^{-1}} for redshifts between \numrange{0.001}{0.1} and steepens to values of \SI{2e46}{erg\,count^{-1}} at $z=2$.
Replacing $\lambda_{\textnormal{RF}}$ in Eq. \ref{equ:flulim} with $\lambda_{\textnormal{RL}}$ yields the cluster luminosity limit.
Then, the cluster temperature and mass are updated according to the \citet{2016A&A...592A...3G} temperature-luminosity,
%  where both, temperature and core-included luminosity, are calculated within \SI{300}{\kilo\parsec} and Lx is extrapolated to L500
\begin{equation}
kT_{\textnormal{lim}} = 
\SI[exponent-product = \cdot]{3}{\keV}
\left[
\frac{L_{500,\textnormal{lim}}}{\SI[exponent-product = \cdot]{3e43}{erg\,s^{-1}}}
\left(\frac{E(z)^{-1.64}}{0.71}\right)
\right]^{\frac{1}{2.63}}
\label{equ:kTlim}
,\end{equation}
and the \citet{2016A&A...592A...4L} mass-temperature scaling relation
\begin{equation}
M_{500,\textnormal{lim}} = 
\frac{\SI{e13.56}{\ensuremath{\textnormal{M}_{\odot}}}}{E(z)}
\left(\frac{kT_{\textnormal{lim}}}{\SI{}{\keV}}\right)^{1.69}.
\label{equ:m500lim}
\end{equation}
With these updated temperature and mass estimates, the procedure starts over and iterates until the change in mass is lower than 0.1\%.
As outlined above, we calculated the different selection thresholds for a step-function-like cluster detection (see Sect. \ref{sec:selcrit}) with 40 and 80 counts on the \sscale and \lscale scale, respectively. The flux and luminosity limit of the two angular scales in fields with different exposures are shown in Fig. \ref{fig:flux_z_limit_all} and Fig. \ref{fig:lum_z_limit_all}, respectively.
Figure \ref{fig:m500_z_limit_all} shows the analytical cluster mass and overdensity radius limit as a function of redshift. The \lscale scale has a lower sensitivity at higher redshift because the area is larger, but it performs better at lower redshift than the \sscale scale. This is promising for galaxy group studies with \erosita, assuming that the considered scaling relations hold at these low masses.
The core radius limit as a function of flux is shown in Fig. \ref{fig:rc_flux_counts_limit_all}. The optimal core radius to detect clusters is approximately \SI{1}{\arcmin}. For a smaller extent, the flux threshold increases because the surface brightness profiles decline faster, such that there are fewer counts in the outskirts. For a larger extent, the flat inner core of the beta model profile causes more photons to lie beyond \SI{4}{\arcmin} and \SI{16}{\arcmin}. This also causes the crossing of the scales around \SI{2}{\arcmin}. As expected, the flux threshold decreases with increasing net exposure time.
Figure \ref{fig:cntstot_z_limit} shows the total count limit of clusters on the two considered angular scales as a function of redshift. Toward low redshift, increasingly larger statistics are required to detect a cluster because the angular extent increases. This emphasises the challenge for \erosita to securely detect very nearby extended sources.

%\subsection{Number of clusters}

\begin{figure}
\centering
\resizebox{\hsize}{!}{\includegraphics{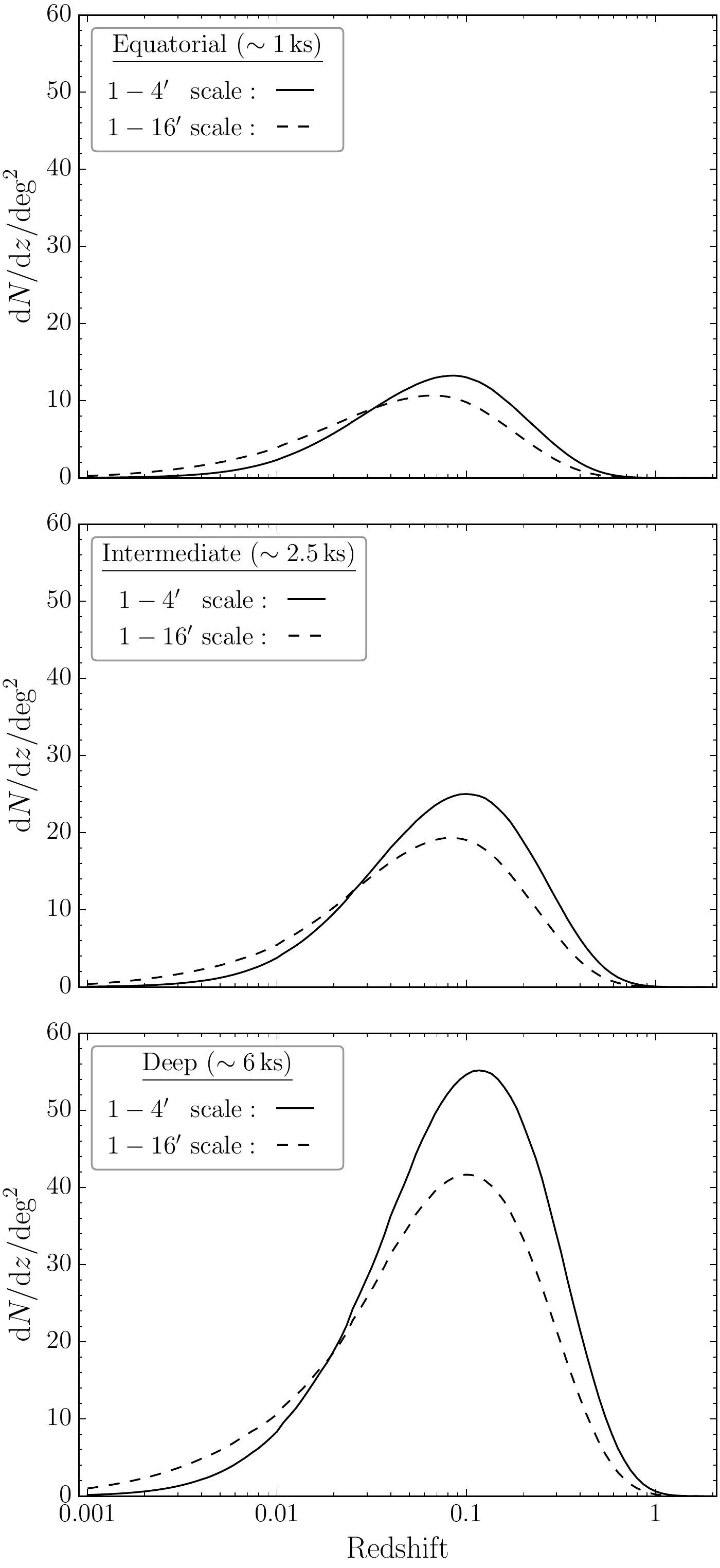}}
%\resizebox{11cm}{!}{\includegraphics{dNdzdeg2_z_limit_all.pdf}}
\caption{Differential number of galaxy clusters per square degree as a function of redshift for the three final \erosita survey fields and the two considered radial scales.
}
\label{fig:dNdzdeg2}
\end{figure}

We used the Python packages COLOSSUS \citep{2018ApJS..239...35D} and Astropy \citep{2013A&A...558A..33A,2018AJ....156..123A} to calculate the differential number of galaxy clusters per square degree at a given redshift by integrating the cluster mass function ($\dd{}n / \dd{}M$, \citealt{2008ApJ...688..709T}) in units of \SI{}{Mpc^{-3}}  multiplied by the differential comoving volume ($\dd{}V / \dd{}z$) in units of \SI{}{Mpc^{3}/deg^{2}} over mass,
\begin{equation}
\frac{\dd{}N}{\dd{}z \,\, \textnormal{deg}^{2}} =
\int_{M_{\textnormal{lim}}(z)}^{M_{\textnormal{max}}}
\frac{\dd{}n}{\dd{}M} \frac{\dd{}V}{\dd{}z} \dd{}M.
\label{equ:diffnumcnts}
\end{equation}
The lower integration limit, $M_{\textnormal{lim}}(z)$, corresponds to the cluster mass limit at the corresponding redshift, and we set the upper limit, $M_{\textnormal{max}}$, to \SI{e16}{\ensuremath{\textnormal{M}_{\odot}}}, above which the contribution of the mass function to the integral is negligible. Figure \ref{fig:dNdzdeg2} shows the differential number of galaxy clusters per square degree as a function of redshift for the three final \erosita  survey fields.
We computed the total number of clusters in a given survey area $A_{\textnormal{s}}$
detected by \erosita according to
\begin{equation}
N = A_{\textnormal{s}}
\int_{M_{\textnormal{lim}}(z)}^{M_{\textnormal{max}}}
\int_{0}^{z_{\textnormal{max}}}
\frac{\dd{}n}{\dd{}M} \frac{\dd{}V}{\dd{}z} \dd{}M \dd{}z.
\label{equ:numcnts}
\end{equation}
For the performance verification (PV) phase of \erosita, a program to reach the average equatorial depth of the final survey on a smaller patch of the sky is planned, the \erosita Final Equatorial-Depth Survey (eFEDS). This will demonstrate the survey capabilities of eROSITA and will allow us to calibrate the scaling relations of galaxy clusters. When we assume an upper redshift limit of $z_{\textnormal{max}}=2$, \SI{3}{\kilo\second} net exposure, and a survey area of 180 square degree, the analytical expectation is to detect approximately 625 clusters using the proposed detection scheme. An in-depth cosmological forecast for galaxy cluster observations with \erosita is left to a future study.

%

%
%%
%%% Section
\section{Simulated field}
\label{sec:simexample}
% eSASSusers_180416

This section demonstrates the performance of the source detection based on wavelet decomposition and characterization on a simulated equatorial \erosita survey field. It serves as an exemplification of the method, and the final adjustments and fine-tuning of the pipeline need to be made on real \erosita data. We simulated the field as described in Sect. \ref{sec:simulation} and processed the output of the simulator using a preliminary version of the \esass package (User release of 2018 April 20).

\subsection{Selection function of extended sources}
\label{subsec:sfextsources}

\begin{figure}
\centering
\resizebox{\hsize}{!}{\includegraphics{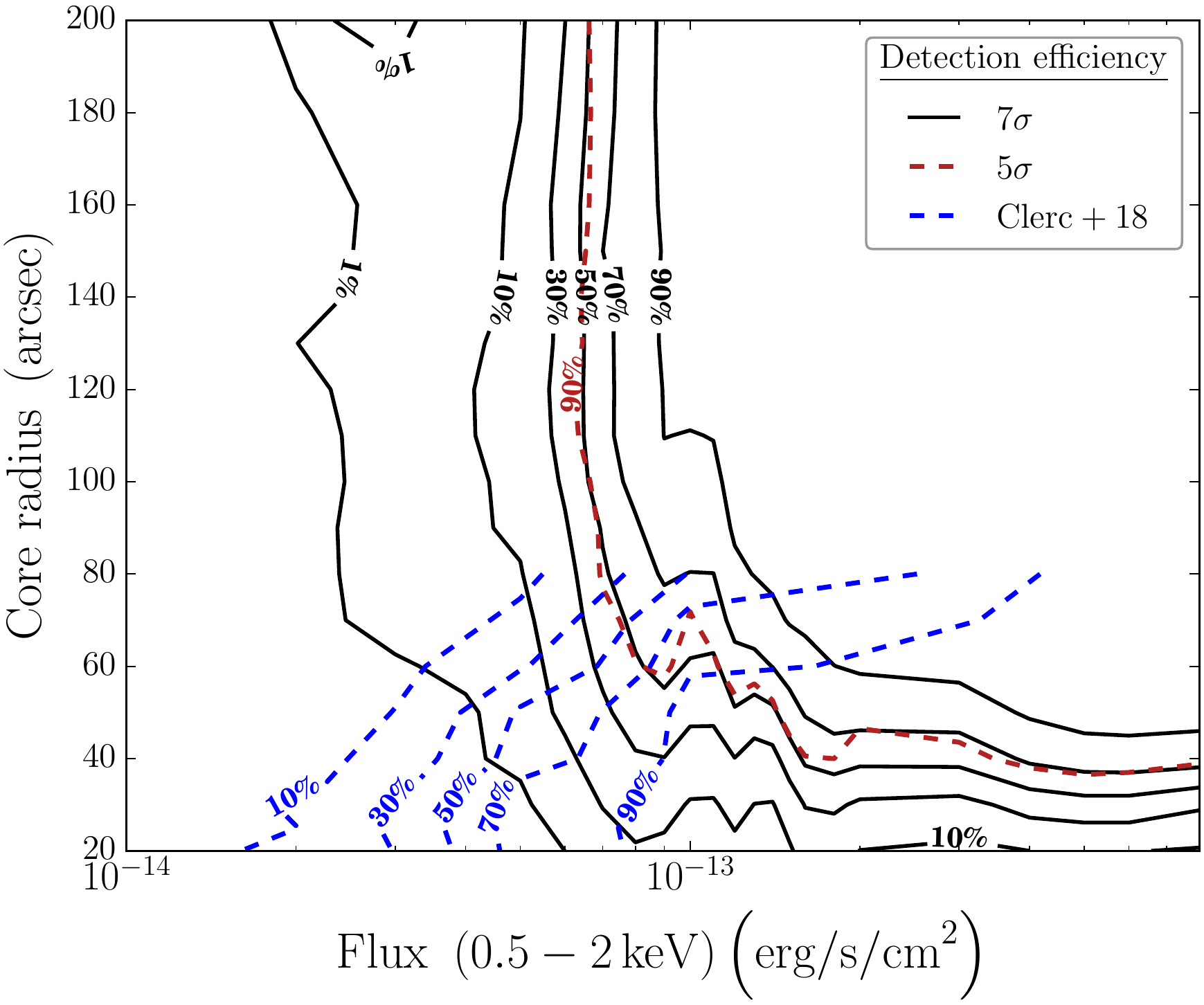}}
\caption{
Extended source detection efficiency of our maximally clean ($7\sigma$ threshold, black contours), our $5\sigma$ threshold (brown contour), and the \citet{2018A&A...617A..92C} threshold (blue contours) in the core radius vs. input flux plain for an equatorial \erosita survey field of approximately \SI{1}{\kilo\second} exposure.
}
\label{fig:flux_rc_contour_wvhs}
\end{figure}

The determination of X-ray survey extended source catalogs and the corresponding selection functions is a trade-off between completeness and purity. The completeness describes the fraction of clusters as function of mass and redshift. Determining it requires an accurate galaxy cluster model because the extended source detection probability depends on the cluster shape. The purity characterizes the contamination of the final sample and requires realistic synthetic simulations. Contamination occurs as a result of point sources that are misclassified as extended or detections that cannot be associated with any input source within a given search radius (spurious detections). We simulated clusters on a predefined spatial grid with a source density such that the emission from neighboring sources did not overlap. This prevented source confusion. The source detection is primarily on the \sscale scale, and we cross-matched extended sources within this typical detection scale of \SI{4}{\arcmin} to the input catalogs. This radius is much smaller than the grid size and slightly larger than the maximum simulated core radius of $3.3^{\prime}$.
We show the maximally clean (\ie $7\sigma$ threshold) extended source detection efficiency in the final equatorial survey field as a function of core radius and input flux in Fig. \ref{fig:flux_rc_contour_wvhs}. Similar to the wavelet decomposition techniques of \citet{1998ApJ...502..558V} and \citet{2007ApJS..172..561B}, our method requires larger photon statistics on compact sources to reduce point-source contamination.
The deficiency of detecting compact objects is the topic of a future study, which relates the angular size to physical scales of the galaxy cluster. A study of the trade-off between cool-core bias and detection efficiency is also deferred to a future work.
The question of how to clean the PED scales is still open. One possibility is to perform a blind analysis by feeding the maximally sensitive source candidate list into the \esass ML fitting routine. For each candidate, a set of source parameters (position, count rate, and extent) was determined by fitting a PSF-convolved $\beta$-model to the spatial distribution of the source counts. The final extended source catalog was compiled by exploring the output parameter space (detection likelihood, extent parameter, and extent likelihood) and by determining appropriate classification thresholds, for instance, to distinguish point-like and extended sources or reduce contamination. This resembles the approach used in \citet{2018A&A...617A..92C} to characterize extended sources that are detected by a sliding-cell algorithm, which scans the X-ray image with a sliding square box of different sizes and weights the counts in the detection box with a $\beta$-model kernel.
This method is a modified version of a sliding-cell and ML fitting adapted for the \xmm Science Analysis Software. \citet{2001A&A...370..689V} compared the performances of several source detection algorithms and found serious drawbacks of this method for the analysis of extended sources because a relatively large number of spurious detections are made and extended sources are split.
The sliding-cell method has a high detection rate of sources with small angular extent ($\lesssim$60$^{\prime\prime}$) at the cost of higher contamination in the ML characterization. When we assume that the detection comes from similar angular scales, the region with most of the misclassified AGNs is excluded when we apply our maximally clean threshold of 7$\sigma$ (extension likelihood of approximately 50) and the extent cut of $60^{\prime\prime}$ to Fig. 9 of \citet{2018A&A...617A..92C}.
Our detection algorithm naturally excludes this highly contaminated region and does not require tuning of extended-source parameters like in the classical wavelet or sliding-cell approach.
Thus, both detection methods can be used complementary or individually to determine discrepancies in the recovered cosmological parameters.
Above $60^{\prime\prime}$, the detection probability stays roughly constant for clusters with larger core radius and does not decrease for clusters up to $200^{\prime\prime}$ because the cluster fluxes are spread over a larger area. Thus, our detection algorithm outperforms the sliding-cell plus ML characterization routine (see Fig. \ref{fig:flux_rc_contour_wvhs}) for large extended sources above approximately 80$^{\prime\prime}$.
In Fig. \ref{fig:flux_rc_contour_wvhs} we also show the 90\% completeness level of the 5$\sigma$ detection threshold. This threshold corresponds to a similar number of detected clusters per square degree between the sliding-cell plus ML characterization algorithm and our method (see below). At the expense of purity, the sliding-cell method is more sensitive for extended sources with core radii smaller than approximately $40^{\prime\prime}$, which correspond to clusters with $r_{500}$ values below \SI{2}{\arcmin}. For the \erosita survey, this gain in sensitivity is a minor effect because the flux of these objects is expected to be close to zero.
Our proposed scheme shows an improvement in detection for flat sources, which are considered as background in other techniques. A more realistic treatment of cluster shapes requires a library of real cluster images, also to properly scale the cool-core emission. This is left for a future study. The classical wavelet approach for \erosita source detection is under development, and we can only compare to the existing study based on the sliding-cell algorithm. The main difference is a change in input list because it also requires an ML characterization.

\begin{figure}
\centering
\resizebox{\hsize}{!}{\includegraphics{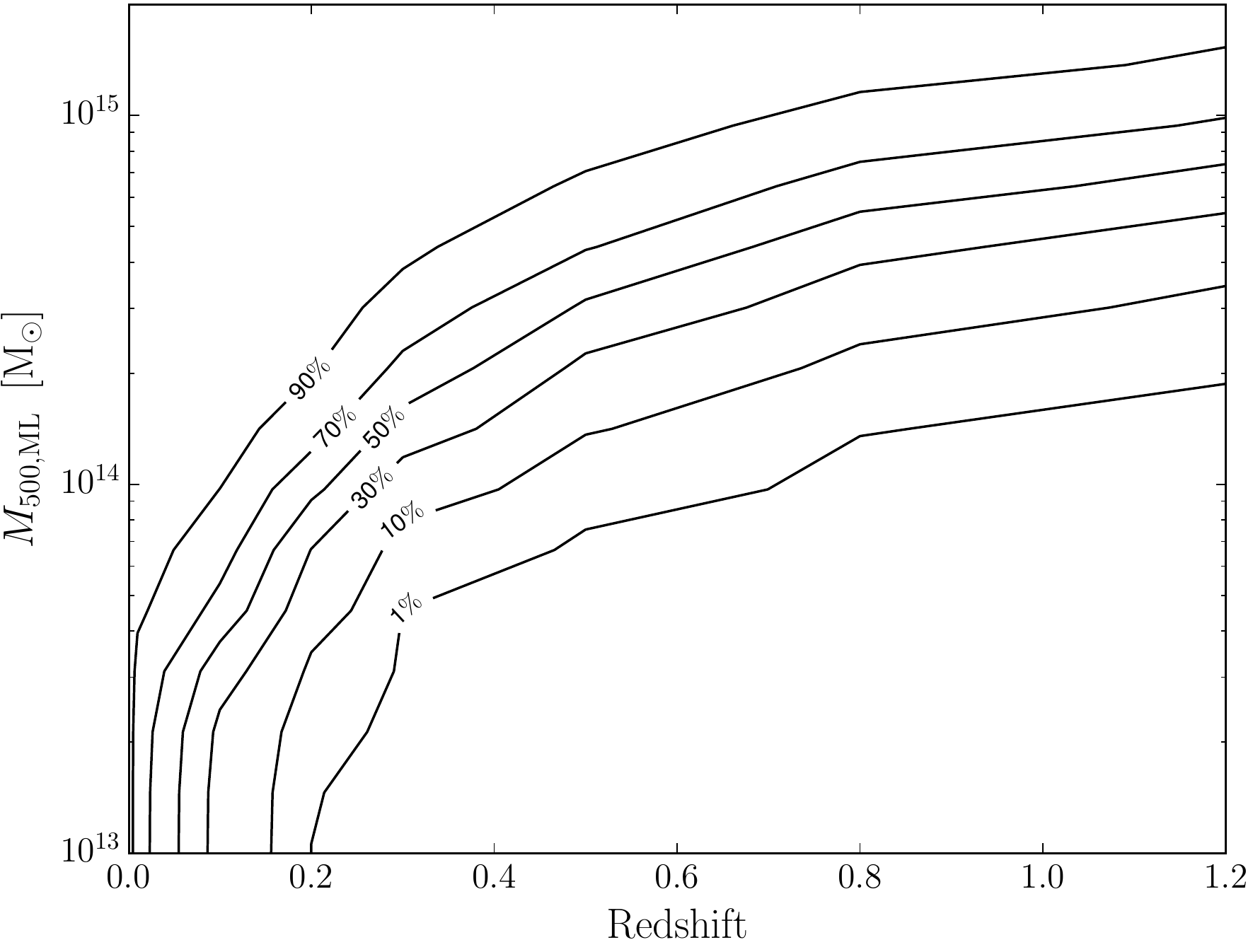}}
\caption{Maximally clean ($7\sigma$) extended source detection efficiency (black contours)  in the mass vs. redshift plain for an equatorial \erosita survey field of approximately \SI{1}{\kilo\second} exposure.
}
\label{fig:z_m500_contour_wvhs}
\end{figure}

Similar to the description in Sect. \ref{sec:theo_prediction}, we used the input temperature to convert the input flux into a luminosity and also used the XXL scaling relations to calculate the galaxy cluster mass, $M_{500,\textnormal{ML}}$. The extended source detection efficiency as a function of mass and redshift is shown in Fig. \ref{fig:z_m500_contour_wvhs}.  The increasing apparent size toward low redshift causes a drop in the detection efficiency.

\begin{figure}
\centering
\resizebox{\hsize}{!}{\includegraphics{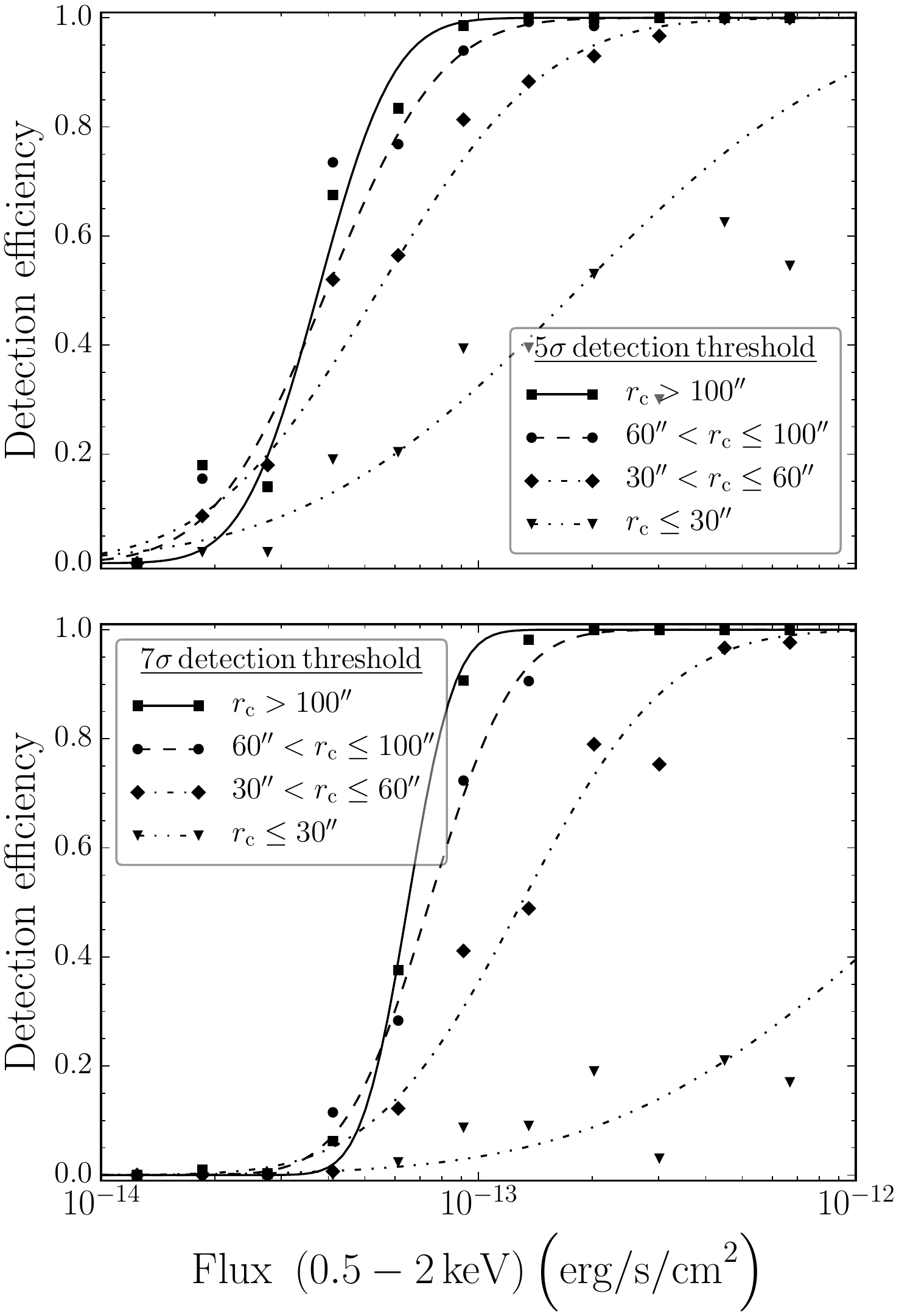}}
\caption{Detection efficiency as a function of input flux for four core radii bins of the $5\sigma$ (upper panel) and $7\sigma$ (lower panel) detection thresholds. Lines correspond to the model expectation of Eq. \ref{equ:selection_analytical} for core radii of 25$^{\prime\prime}$, 50$^{\prime\prime}$, 85$^{\prime\prime}$, and 150.
}
\label{fig:deteff_flux_2090}
\end{figure}

\begin{figure}
\centering
\resizebox{\hsize}{!}{\includegraphics{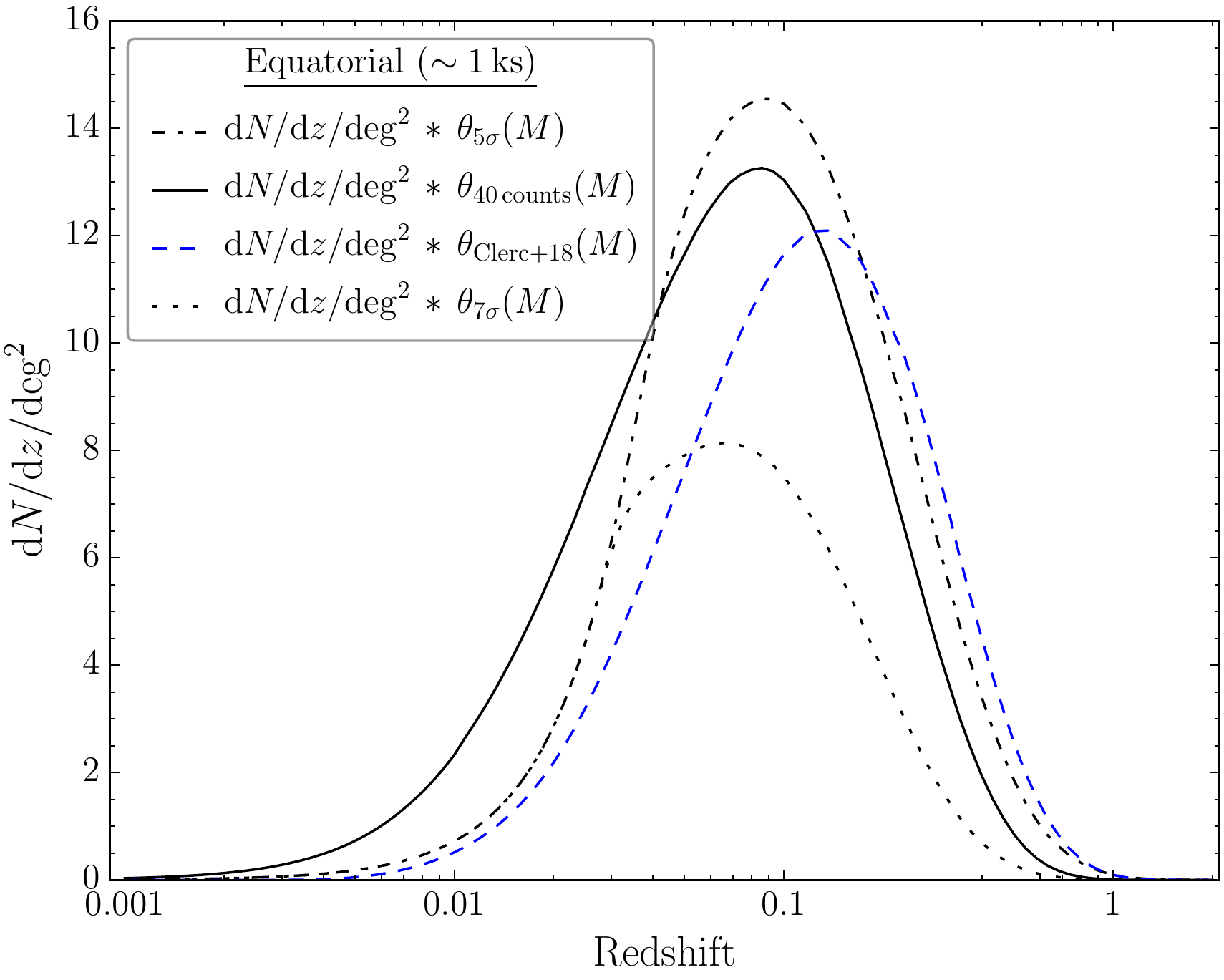}}
\caption{Expected differential number of galaxy clusters per square degree as a function of redshift for the final equatorial \erosita survey field with different selection functions folded in. The solid line serves as comparison and show the 40 aperture count selection on the \sscale scale of Fig. \ref{fig:dNdzdeg2}.
}
\label{fig:dNdzdeg2_sff}
\end{figure}

We folded the $5\sigma$ and $7\sigma$ selection on the \sscale scale (Figs. \ref{fig:flux_rc_contour_wvhs} and \ref{fig:z_m500_contour_wvhs}), as well as the sliding-cell selection \citep[][Appendix A]{2018A&A...617A..92C}, into the calculation of the differential number of clusters per square degree by multiplying the mass function in Eq. \ref{equ:diffnumcnts} with the probability of detecting a cluster of the given mass, that is, the selection function $\theta(M)$,
\begin{equation}
\frac{\dd{}N}{\dd{}z \,\, \textnormal{deg}^{2}} =
\int_{M_{\textnormal{min}}=\SI{e13}{\ensuremath{\textnormal{M}_{\odot}}}}^{M_{\textnormal{max}}=\SI{e16}{\ensuremath{\textnormal{M}_{\odot}}}}
\frac{\dd{}n}{\dd{}M} \frac{\dd{}V}{\dd{}z} \theta(M) \dd{}M. 
\label{equ:diffnumcnts_sf}
\end{equation}
In practice, we analytically parameterized the selection as a function of core radius and flux. The overall functional form of the detection efficiency is described by an error function, which was scaled to range between zero and one. The overall shape of the error function is defined by its argument. Compared to \citet{2018A&A...617A..92C}, we required a more complex functional form of the argument because it needs to describe a change in slope for different core radii in addition to an offset in flux for different core radius values. 
The goal is to find a functional form that is as simple as possible but still accounts for these observed features. The functional form of the argument is found by iteratively adding more complexity to it until the detection efficiency is described well. Then, the free parameters are optimized using a Markov chain Monte Carlo posterior sampling technique. Therefore, the functional form has no physical motivation. To improve the iterative finding of the functional form, we reduced the dynamical range of the core radius and flux by taking the logarithm and subtracted the corresponding means to rescale the offsets.
The selection function is described best according to
\begin{equation}
\label{equ:selection_analytical}
\begin{aligned}
\theta(R, F) &= 0.5 + 0.5 \cdot \textnormal{erf} \left(
a +
b \cdot R +
c \cdot F +
\exp \left( d \cdot R \right) \right)\\
R &= \log \left( r_{\textnormal{c}} / \left[\textnormal{arcsec} \right] \right) - 2 \\
F &= \log \left( \textnormal{flux} / \left[\SI{}{erg\,s^{-1}\,cm^{-2}}\right] \right) + 13.
\end{aligned}
\end{equation}
The parameters $a$, $b$, $c$, and $d$ depend on the detection threshold.
We show the models and their parameters in Fig. \ref{fig:deteff_flux_2090} and Table \ref{tab:analytic_params} for the $5\sigma$ and $7\sigma$ thresholds. These simple models cannot capture the complexity of the selection, but they provide a good estimate of the detection efficiency.
\begin{table}
\caption{Best-fit parameters of the analytic selection function (Eq. \ref{equ:selection_analytical}) for the $5\sigma$ and $7\sigma$ detection thresholds.}
\label{tab:analytic_params}
\centering
\begin{tabular}{c c c c c}
\hline\hline
Detection threshold & $a$ & $b$ & $c$ & $d$ \\
\hline
5$\sigma$ & 1.40 & 2.86 & 3.35 & 1.66 \\
7$\sigma$ & 0.77 & 3.42 & 4.78 & 2.43 \\
\hline
\end{tabular}
\end{table}
The impact of the different selection functions on the differential number counts is shown in Fig. \ref{fig:dNdzdeg2_sff}.
The expected number of galaxy clusters per square degree for the \citet{2018A&A...617A..92C} and the $5\sigma$ selection is approximately 4.2. At the cost of reduced purity, high-redshift clusters are detected more efficiently by the sliding-cell algorithm plus ML fitting technique, while the method based on wavelet decomposition performs much better in detecting the local population, that is, in particular galaxy groups. The $7\sigma$ selection reduces the contamination by more than two orders of magnitude (see Sect. \ref{subsec:selection}), but the number of detected clusters per square degree is, with approximately 1.7, more than halved.

We require better knowledge of how the background behaves in reality to securely forecast the detection of very extended low-redshift objects for which the core radius limits are larger than $200^{\prime\prime}$.
The uncertainty on small scales is dominated by the unknown shape of the survey PSF. The \erosita PSF does not vary much over the \erosita field of view compared to other X-ray instruments like \chandra and is, to first approximation, constant in survey mode. An interesting planned implementation for our proposed method is therefore subtracting point sources using a precise PSF model in the ML fitting routine.

\subsection{Selection}
\label{subsec:selection}

\begin{figure}
\centering
\resizebox{\hsize}{!}{\includegraphics{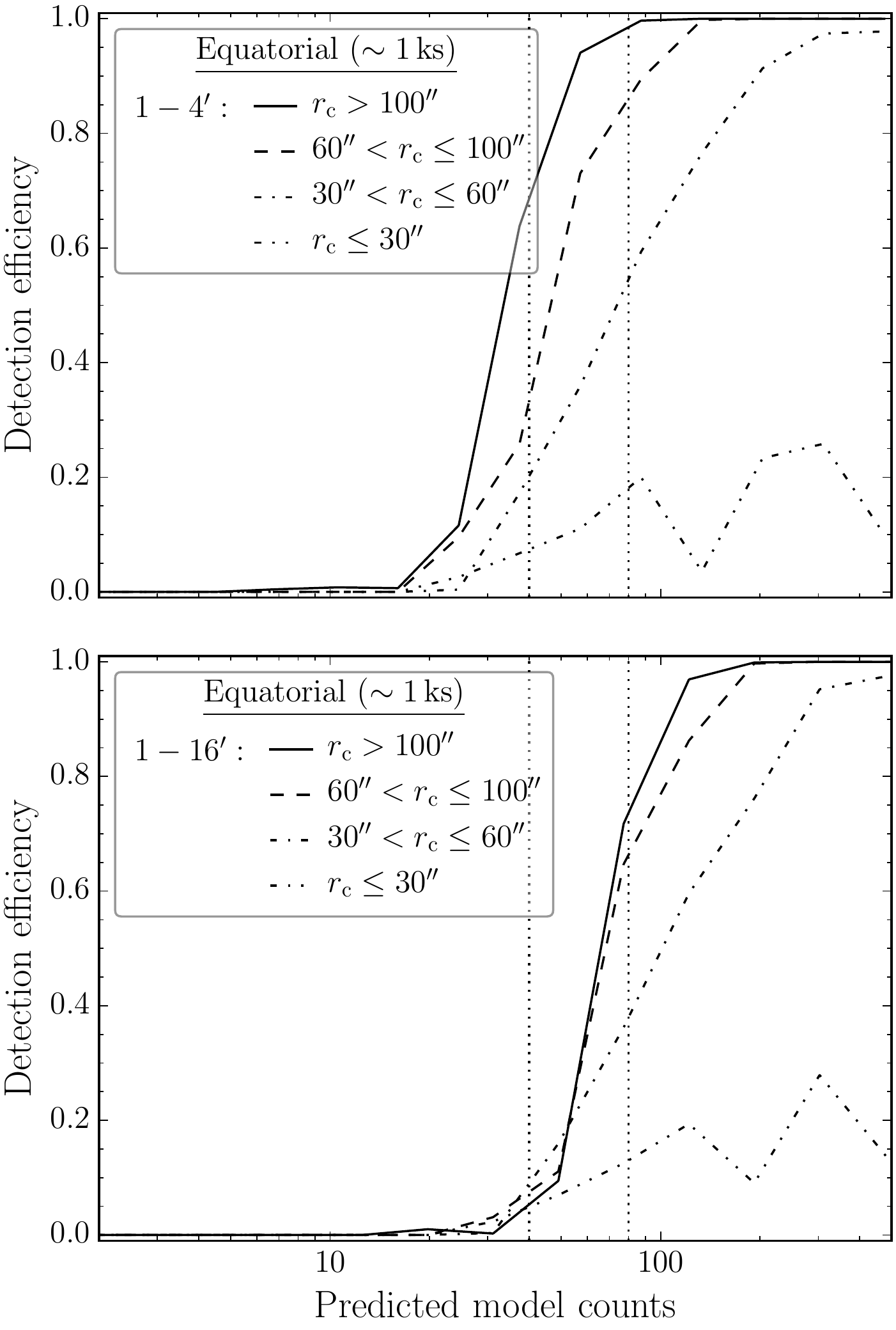}}
\caption{Detection efficiency as a function of predicted model counts on the \sscale (upper panel) and \lscale (lower panel) radial scale for four core radius bins. The dotted vertical lines correspond to 40 and 80 aperture counts, respectively.
}
\label{fig:deteff}
\end{figure}

\begin{figure}
\centering
\resizebox{\hsize}{!}{\includegraphics{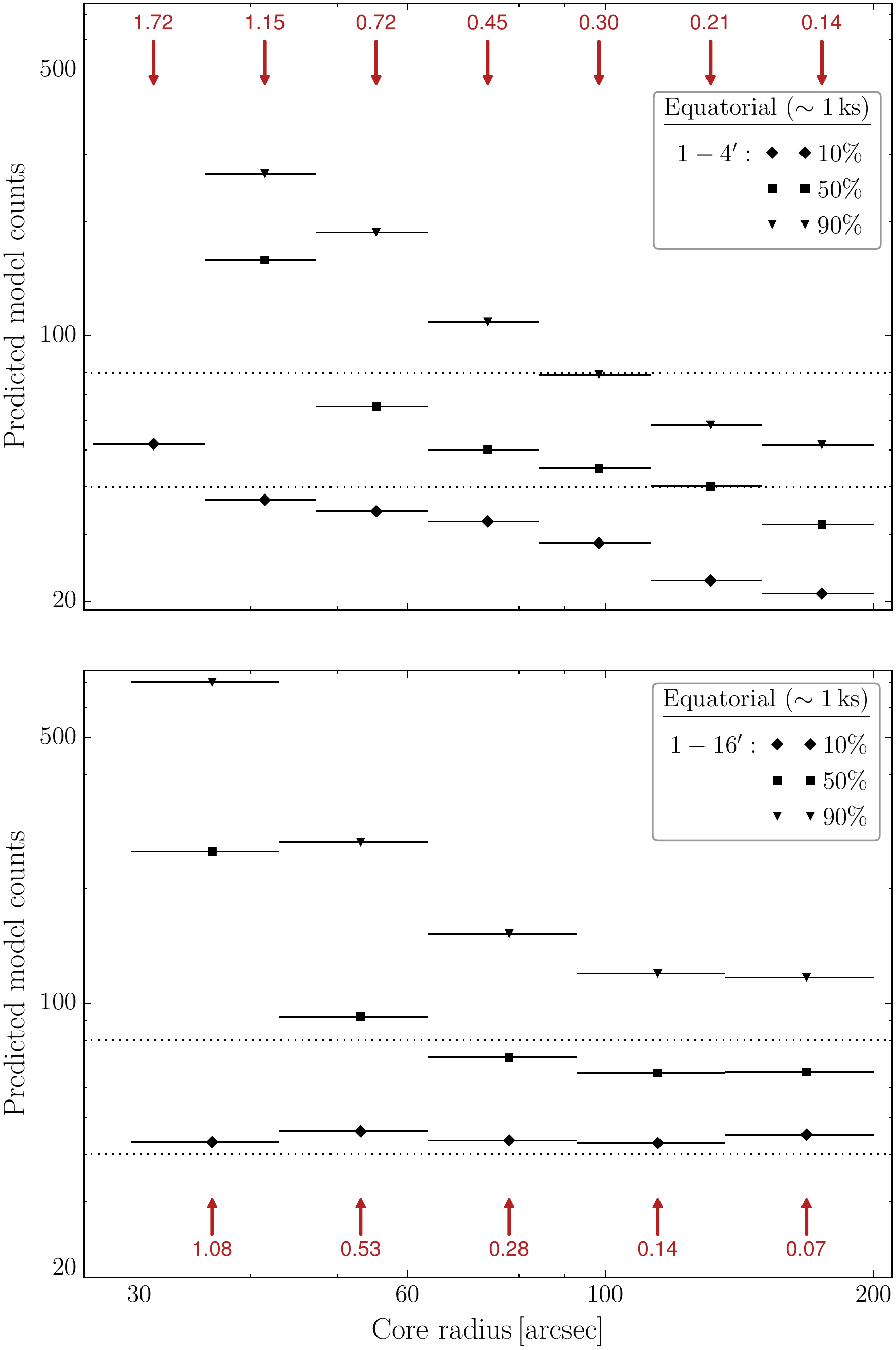}}
\caption{Predicted model counts for a 10\%, 50\%, and 90\% detection efficiency as a function of core radius on the \sscale (upper panel) and \lscale (lower panel) radial scale. The brown values indicate the ratio of the model counts in the \cscale and the corresponding angular scale of the individual core radius bin. The dotted horizontal lines correspond to 40 and 80 aperture counts, respectively.
}
\label{fig:mcounts_rc}
\end{figure}

\begin{figure}
\centering
\resizebox{\hsize}{!}{\includegraphics{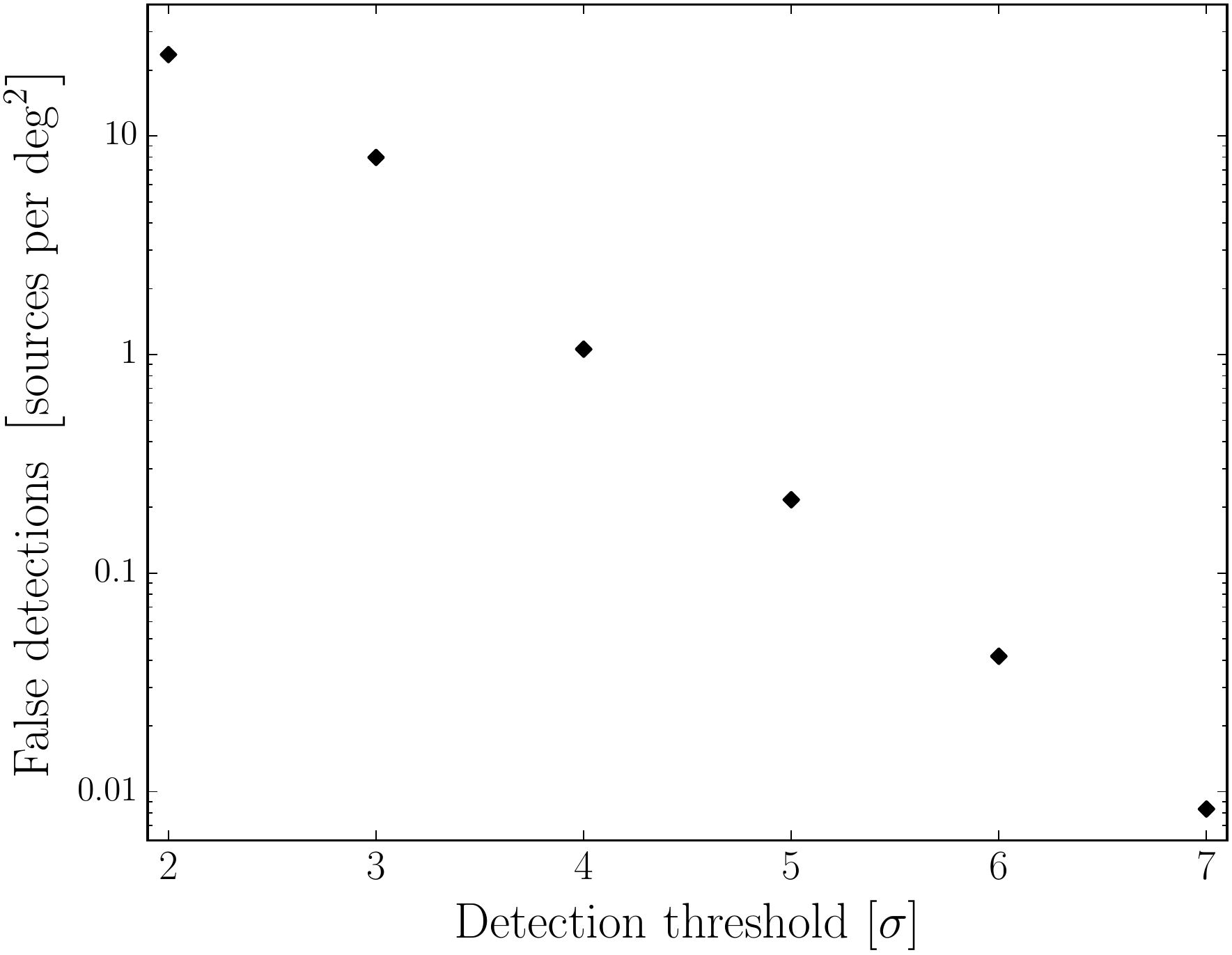}}
\caption{Number density of spurious and misclassified extended sources as a function of detection threshold for an \erosita survey exposure of approximately \SI{1}{\kilo\second}.
}
\label{fig:purity_threshold}
\end{figure}

\begin{figure*}
\centering
\resizebox{\hsize}{!}{\includegraphics{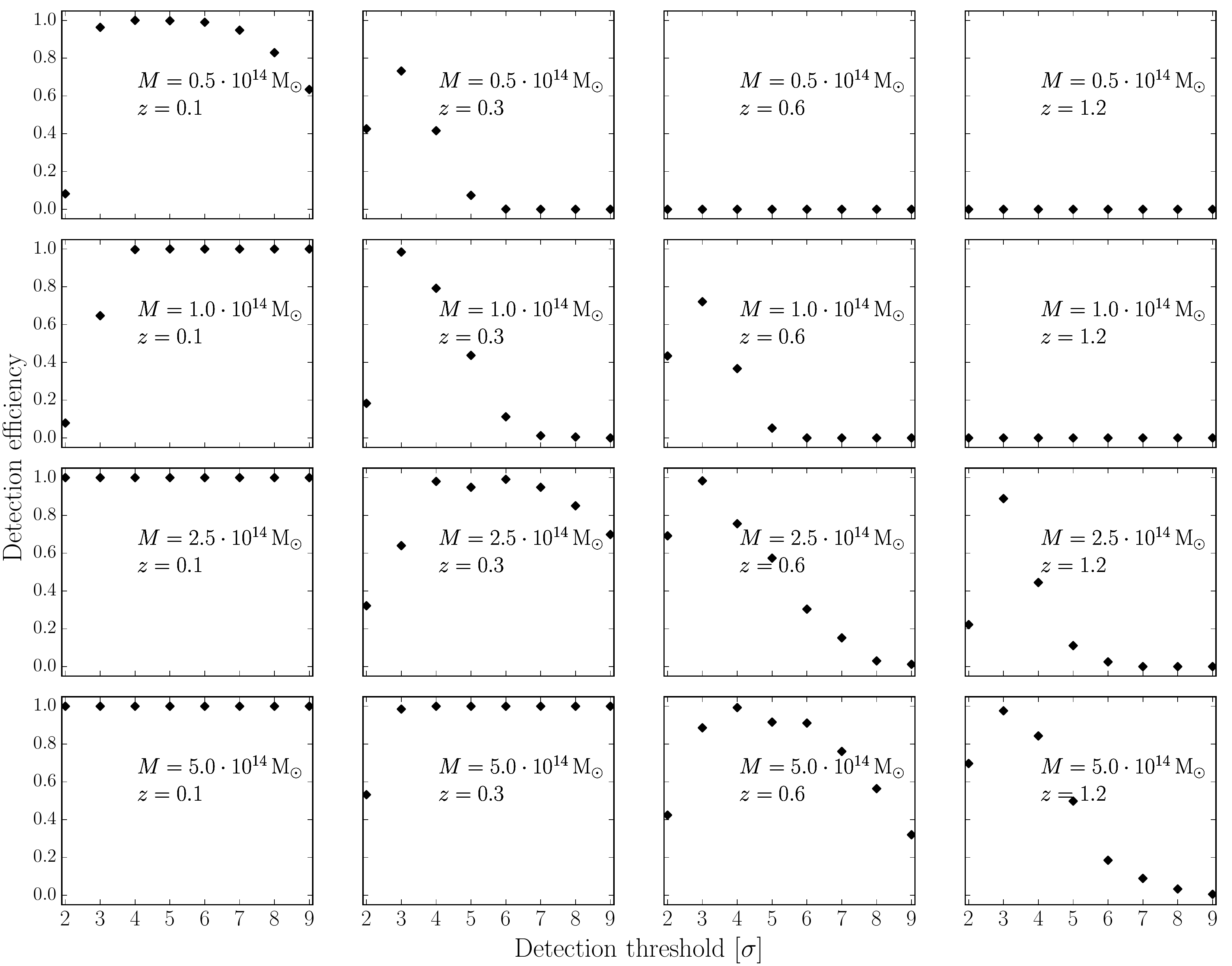}}
\caption{Detection efficiency of extended sources as a function of detection threshold for snapshots in redshift and mass for an \erosita survey exposure of approximately \SI{1}{\kilo\second}.
}
\label{fig:compl_threshold}
\end{figure*}

We address the question how well the detection through cluster outskirts resembles a favored step-function-like selection. Figure \ref{fig:deteff} shows the detection efficiency
on two angular scales and different core radius bins as a function of predicted model 
counts, which are independent of PSF effects. For a given number of predicted counts, clusters with larger extent are detected more efficiently. In other words, even with a larger number of predicted counts, clusters with smaller extent are harder to detect.
The interesting finding that gradually increasing counts toward smaller core radii are required is summarized in Fig. \ref{fig:mcounts_rc}, showing the predicted model counts for a given detection efficiency as a function of core radius. In addition, it shows values of the model count ratio on the \cscale over the \sscale and \lscale radial range, respectively. This emphasizes that for a given detection efficiency, the required counts in the outskirts increase with increasing inner-to-outer counts ratios. Considering an additional contribution of AGN in cluster centers, this is particular challenging for clusters above a redshift of \num{0.6}, where simulations indicate that the distribution of the ratio becomes broader and exhibits a significant fraction larger than two \citep{2018MNRAS.481.2213B}. These findings motivate the estimation of contamination due to bright sources and due to low photon statistics separately because the flux distribution of faint sources is different from that of bright sources. We studied these two effects by creating two extended source catalogs, setting the detection thresholds of cataloging to 4$\sigma$ and 7$\sigma$ for the maximally sensitive and maximally clean selection, respectively.
The number of false detections as a function of detection threshold is shown in Fig. \ref{fig:purity_threshold}. We obtain close to $1.1$ and $0.008$ spurious or misclassified extended sources per square degree in equatorial fields for the 4$\sigma$ and 7$\sigma$ detection thresholds, respectively. Detection thresholds greater than 7$\sigma$ show zero contamination but also a lower detection efficiency in regimes of low photon statistics.
The extended source detection efficiency as a function of detection threshold is exemplified, showing mass and redshift dependencies in Fig. \ref{fig:compl_threshold}. In several cases, the efficiency for the 2$\sigma$ threshold drops because the algorithm keeps so much structure that the extracted sources cannot be associated with the correct input within the given matching radius.

%

%
%%
%%% Section 
\section{Summary and conclusions}
\label{sec:summary}

Large-area X-ray cluster surveys are powerful tools for deriving cosmological parameters when the selection effects are well understood. We proposed and characterized an algorithm based on a wavelet decomposition to detect extended source for the upcoming \erosita mission. This technique produces well-defined cluster catalogs with simple selection functions. We detect clusters by their large-scale emission, which minimizes the predominant impact of excess cool-core emission. Our main result is that progressively more counts are required with decreasing cluster extent to achieve a specific detection efficiency. In addition, our analytical calculation shows that an increasing number of total counts toward low redshift is required, meaning a larger angular extent, to detect clusters as extended sources. These two findings disgree with the assumption that a fixed minimum number of total photons are necessary to identify clusters \citep[\eg][]{2012MNRAS.422...44P,2014A&A...567A..65B}.
We predict redshift-dependent cluster observables and mass limits for an equatorial, intermediate, and deep final \erosita survey field by assuming a minimum number of 40 and 80 counts to identify a cluster on a \sscale and \lscale angular scale, respectively. The counts in the cluster outskirts define an easy-to-measure observable, and applying a minimum photon threshold provides a selection that approximately resembles a step function.
We tested the performance of our detection scheme through Monte Carlo simulations of a final equatorial \erosita survey field of approximately \SI{1}{\kilo\second} net exposure time. Our maximally clean detection method requires larger photon statistics on objects with core radii smaller than $60^{\prime\prime}$ to minimize point-source contamination and has an approximately 90\% detection efficiency at input fluxes of \SI{e-13}{erg\,s^{-1}\,cm^{-2}} for clusters with larger extent. This is complementary to the sliding-cell algorithm plus ML fitting technique that is currently implemented as default in \esass, which shows a drop in detection efficiency at this flux for clusters with core radii larger than $60^{\prime\prime}$ \citep{2018A&A...617A..92C}.
We note that this blind analysis approach increases the contamination of the final catalog by misclassified AGNs and spurious extended sources.
At a similar level of completeness, our catalogs are approximately 2.5 times purer than the current \esass default.
Our performance results are limited because we worked with pre-flight assumptions of instrumental and astrophysical characteristics. The proposed pipeline has the advantage that the final tuning, that is, the point-source model training due to a different in-orbit PSF or the optimized selection of the detection thresholds, is easy to implement, robust, and can be achieved very fast during the PV phase. An in-flight calibration of the pipeline below 5\% is expected to keep the loss of clusters through central AGN contributions below 1\%.

%
%%
%%% Acknowledgments
\section*{Acknowledgments}
This research was made possible by the International Max Planck Research School on Astrophysics at the Ludwig-Maximilians University Munich, as well as their funding received from the Max-Planck Society.

%
%%
%%% References
\bibliographystyle{aa}
\bibliography{detection_outskirts}

\end{document}